\documentclass[9pt,twocolumn,twoside]{osajnl}
\journal{josab} 
\setboolean{shortarticle}{false}

\title{Low-power optical bistability in  $\mathcal{PT}$-symmetric chirped Bragg gratings with four-wave mixing}

\author[1]{S. Sudhakar}
\author[2]{S. Vignesh Raja}
\author[3]{A. Govindarajan}
\author[4,* ]{K. Batri}
\author[3]{M. Lakshmanan}

\affil[1]{Department of Electronics and Communication Engineering, Vel Tech High Tech Dr. Rangarajan Dr. Sakunthala Engineering College, Chennai - 600 062, India}
\affil[2]{Department of Electronics and Communication Engineering, SRM Institute of Science and Technology, Ramapuram Campus, Chennai - 600 087, India}
\affil[3]{Centre for Nonlinear Dynamics, School of Physics, Bharathidasan University, Tiruchirappalli - 620 024, India}
\affil[4]{Department of Electronics and Communication Engineering, PSNA College of Engineering and Technology, Dindigul - 624 622,India}
\affil[*]{Corresponding author: krishnan.batri@gmail.com}

\dates{Compiled \today}

\ociscodes{(140.3490) Lasers, distributed feedback; (060.2420) Fibers, polarization-maintaining;(060.3735) Fiber Bragg gratings.}

\doi{\url{http://dx.doi.org/10.1364/XX.XX.XXXXXX}}

\begin{abstract}
The central theme of this article is the analysis of the usefulness of introducing four-wave mixing or modulation of Kerr nonlinearity in a nonuniform grating structure with gain and loss. To do so, we propose an inhomogeneous system in which the  nonlinearity of  the $\mathcal{PT}$-symmetric grating is modulated. It is proven that the proposed scheme can be fruitful only if the nonlinearity and its associated modulation terms are assumed to be self-defocusing ones. In order to cut down the intensity required for the steering, the sign of chirping is taken to be negative and the wavelength of operating light is considered to be higher than the Bragg wavelength. Alongside these settings, launching the light from the rear end of the device dramatically reduces the critical intensity to a value of 0.015 (approximately) which must be the lowest switching intensity ever reported in the context of nonlinear $\mathcal{PT}$-symmetric gratings. The ramp-like stable states in the broken $\mathcal{PT}$-symmetric regime prevail even in the company of modulation in the nonlinearity index for both of the light incident directions.
\end{abstract}

\setboolean{displaycopyright}{true}

\begin{document}
	\maketitle

	\section{Introduction}

	Fiber Bragg gratings (FBGs) are one of the inevitable building blocks in the modern-day architecture of an all-optical network \cite{erdogan1997fiber,othonos1997fiber,giles1997lightwave}. In the pursuit of realizing an integrable all-optical and low-power FBG switch, researchers have come up with a lot of innovative design solutions \cite{radic1994optical,radic1995theory,yousefi2015all,ping2005bistability}. Among the wide variety of schemes available, the idea of modulation of Kerr nonlinearity \cite{chen2016evolution, chen2015propagation, chen2016propagation, peng2016interaction, zhang2019dynamics} stands out to be one of the important methodologies to produce low-power bistable optical switches \cite{pelinovsky2002stableI,pelinovsky2002stable,winnie2003stable,komissarova2019pt}. The reason for the disappearance of multistable states in these types of systems is that the cubic nonlinearity (Kerr) is compensated by a high value of nonlinearity averaged over the structure. The above kind of system was thoroughly investigated by Pelinovsky \emph{et al.} in the perspective of optical bistability (OB) under two regimes, namely zero (balanced) and non-zero (non-balanced) modulation of Kerr nonlinearity regimes \cite{pelinovsky2002stableI}. The former requires a nonlinear (NL) modulation index ($n_K$) to be kept at a non-zero value but the magnitude of the average value of nonlinear RI ($n_{nl}$) is set to zero. Under this regime, it is possible to achieve true-optical limiting \cite{pelinovsky2002stable}. True optical limiters are those nonlinear devices which favor uniformly stable operating regimes for all pertinent incident intensities and this mechanism is called all-optical limiting \cite{pelinovsky2002stable,pelinovsky2002stableI,phang2013ultrafast}.  But, the other regime (in which both these coefficients are kept at non-zero values) is more interesting for the reason that it brings an additional degree of freedom (in the form of the parameter $n_K$) for the realization of low-power NL applications like switching, memory and signal processing \cite{pelinovsky2002stableI,komissarova2019pt}.  Hence, we explore this regime in this article in the presence of $\mathcal{PT}$-symmetry and chirping nonuniformity. 
	
 Physically, this modulation of Kerr nonlinearity term would have originated because of the complex four-wave mixing (FWM) process happening inside the fiber because of the fundamental interactions among different frequency components \cite{xie2020chip}. 	The literature provides a clear scientific explanation or practical validation in the form of experiments for the origin of the FWM in physically realizable gratings \cite{ko2019enhanced,bencivenga2015four,wu2016nearly,ullah2014observation,che2020kerr,hu2020propagating}. Nonlinear processes inside an optical fiber involving modulation of one or more system parameters are generally referred as parametric processes and in the present work the Kerr nonlinearity parameter is modulated. For such a parametric process, to get a build up in any fiber based device, phase matching is an essential ingredient. The FWM arises as a result of the nonlinear response of bound electrons of a material to the incident optical field \cite{agrawal2001applications}. The interaction between these frequency components generate a new frequency component at the output end. Remarkably, the signal generated by the FWM process also occurs within the photonic bandgap (PBG) (range of wavelengths in which the transmission of incident optical light is inhibited \cite{erdogan1997fiber}) of the grating. As a result, they can be generated at desired frequencies simply by manipulating the system and the signal parameters \cite{agrawal2001applications, xie2020chip}. 

	Presently, the notion of $\mathcal{PT}$-symmetry is reckoned as one of the research areas to be subjected to an intense investigation by the scientific community (See Refs. \cite{el2007theory,el2018non,kottos2010broken,kulishov2005nonreciprocal,ozdemir2019parity,govindarajan2019tailoring,govind2019,sarma2014modulation,longhi2018parity,miri2012bragg,kartashov2014unbreakable} and therein). This notion would have remained as an unrealistic subject in real physical settings forever if not for the fact that the fundamental postulates of $\mathcal{PT}$-symmetry can be translated from quantum mechanics to optics \cite{bender2007making,el2007theory,ruter2010observation}. As far as FBGs are concerned, $\mathcal{PT}$-symmetric structures reveal some atypical light propagation characteristics such as direction-dependent (linear) \cite{lin2011unidirectional,kulishov2005nonreciprocal,huang2014type,raja2020phase,raja2020tailoring,lupu2016tailoring,raja2021n}, nonreciprocal (nonlinear) \cite{raja2019multifaceted,raja2019nonlinear,govindarajan2020tunable}, and unidirectional \cite{lin2011unidirectional} wave transport provided that the mathematical condition $n(z) = n^*(-z)$ for the refractive index is fulfilled by the structure via a precise equilibrium between gain and loss regions. Also, it is proven that  different kinds of nonlinearity including Kerr modulation could essentially suppress the instability in certain $\mathcal{PT}$- symmetric systems with equal amount of gain and loss \cite{huang2014pt, luz2019robust}.

	In the context of uniform $\mathcal{PT}$-symmetric FBGs (PTFBGs), quite a good number of nonlinear models have been proposed so far to realize all-optical switches. To cite a few, time domain and time-independent models of PTFBG with Kerr nonlinearity \cite{liu2014optical}, a combination of third-order nonlinearity with dispersive and saturable gain and loss profiles \cite{phang2013ultrafast,phang2014impact,phang2015versatile}, PTFBGs with modulated Kerr nonlinearity \cite{komissarova2019pt}, and higher-order nonlinearities \cite{raja2019multifaceted} are some of the important works in the literature, as of now. Subsequently, three of the present authors with Mahalingam came up with a theoretical modeling of a low power switch involving a nonlinear  PTFBG with a chirping nonuniformity \cite{raja2019nonlinear}.

	Chirping is a kind of nonuniformity that is intentionally introduced into a grating in order to vary the spatial frequency of the device (rather than keeping it constant as in the case of uniform gratings) when the laser light is launched into it. As mentioned above, the variations in chirping can either increase (positively chirped) or decrease (negatively chirped) the spatial frequency along the propagation direction ($z$) (See Refs. \cite{kim2001effect,maywar1997transfer,maywar1998effect,maywar1998low,ouellette1987dispersion} and therein). In this article, we investigate the optical bistability features exhibited by a chirped PTFBG (CPTFBG) whose nonlinear RI is modulated over the propagation distance, and from here onwards we denote the corresponding term by $n_K$ or call it as modulated Kerr nonlinearity term. In the literature, one can find that the authors have considered positive (self-focusing) and negative values (self-defocusing) for both the cubic nonlinearity ($\gamma_{nl}$) and modulation of Kerr nonlinearity ($\gamma_K$) parameters. Our motivation is to find the best regime among the different operating domains of the system that leads to a low-power switching as discussed in Sec. \ref{Sec:III}\ref{Sec:3a}. Even though the same study (without introducing the detuning and chirping) has been carried out by Kommisorrova \emph{et al.} \cite{komissarova2019pt}, the present work is different from their work due to the following considerations: The current study includes the chirping nonuniformity.
	 Due to the presence of chirping, different combinations of chirping ($C$), detuning ($\delta$), Kerr nonlinearity ($\gamma_{nl}$), and modulation of Kerr nonlinearity ($\gamma_K$) parameters can give rise to different kinds of OB curves and these sign combinations (positive or negative) are likely to have a key role in dictating the spectral span and power required to switch between the on and off states of a bistable curve which was not reported in detail in {\color{blue}Ref} \cite{komissarova2019pt}. The dynamics of a nonuniform PTFBG with its nonlinearity term modulated in the broken regime needs to be depicted in an elaborate fashion and so an individual section (Sec. \ref{Sec:III}\ref{Sec:3b}) is dedicated to this study. The presence of perturbation in the form of modulated Kerr nonlinearity casts doubt upon the nonreciprocal and low power switching dynamics under the reversal of direction of light incidence and hence a separate numerical investigation is required (from this perspective) and the results are shown in Sec. \ref{Sec:IV}.

	\label{sec:I}

	\section{Model}
	\label{sec:II}
	\begin{figure*}[t]
		\centering	\includegraphics[width=1\linewidth]{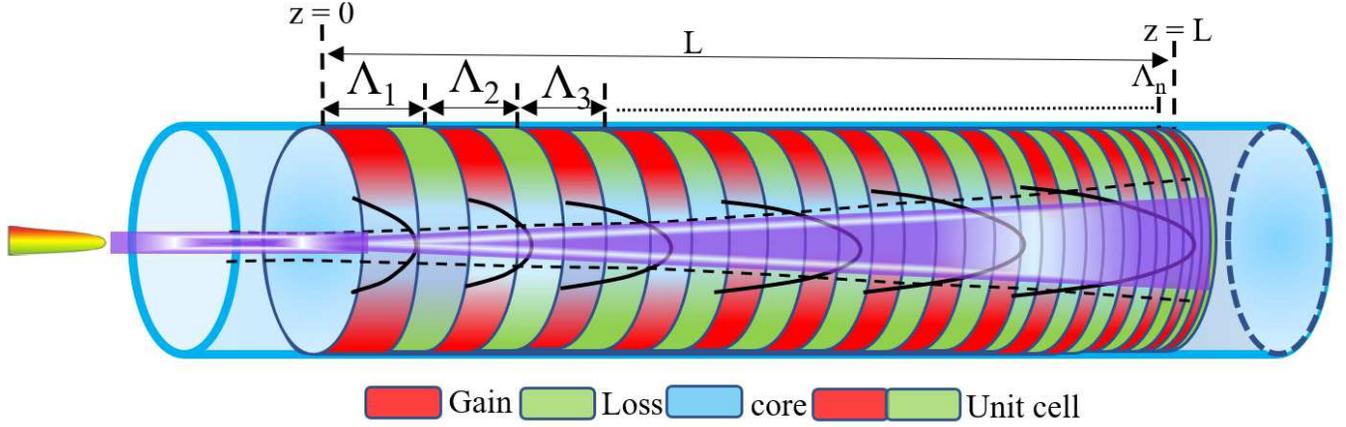}
		\caption{Schematic of a nonlinear $\mathcal{PT}$ -symmetric chirped FBG. The system satisfies $\mathcal{PT}$-symmetric condition in each unit cell featuring alternate regions of gain and loss as indicated in red and green, respectively. The nonlinear interaction of the incident field inside the device  is self-defocusing type.  }
		\label{fig0}
	\end{figure*}

\begin{figure}[t]
	\centering	\includegraphics[width=0.7\linewidth]{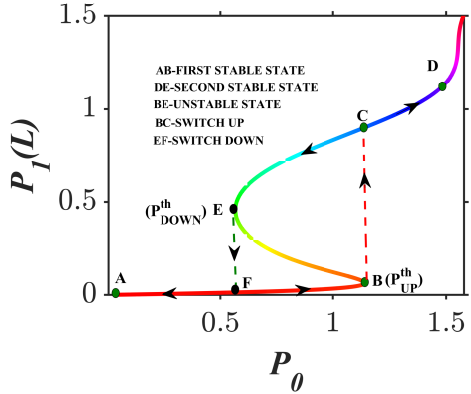}
	\caption{Illustration of a typical OB curve showing various stable, unstable states, and switching intensities. }
	\label{fig0B}
\end{figure}
	Consider a PTFBG of length $L$ whose grating period ($\Lambda$) is chirped [$\Lambda(z)$] along the direction of propagation of light ($z$) as shown in Fig. \ref{fig0}. Modern-day fiber optic experiments use high-power laser sources and so invoking the nonlinearity inside the fiber will not be a tedious task. For our theoretical investigation, we consider a high-power laser source with input intensity $P_0$ that induces a third-order nonlinearity ($n_{nl}$) inside the CPTFBG. Unlike highly nonlinear chalcogenide fibers, conventional silica fibers support only cubic nonlinearity when a continuous wave laser is launched into it. It is assumed that the layered and distributed feedback structure accounts for the modulation of nonlinear RI coefficient ($n_k$) in addition to the modulation of the linear RI terms ($n_{1R}$ and $n_{1I}$).  As the input field is launched into the system, the beam expands as it propagates inside the system due to the self-defocusing effect and the nonlinear interaction is pictorially depicted in the system model shown in Fig. \ref{fig0}.  The RI distribution of the system that describes the combined effects of RI of the background material ($n_0$), modulation of real ($n_{1R}$) and imaginary ($n_{1I}$) parts of RI, third-order nonlinear coefficient ($n_{nl}$), modulation of the third-order nonlinear coefficient ($n_K$) is given by
	\begin{gather}
	\nonumber n(z)=n_0+n_{1R}cos(2 \pi z/\Lambda)+in_{1I}sin(2 \pi z/\Lambda)+n_{nl}|E|^2\\+n_{K}|E|^2 cos(2 \pi z/\Lambda)
	\label{Eq: Norm1}
	\end{gather}
	The optical field ($E(z)$) launched into the CPTFBG can be described in mathematical terms as the sum of the envelope of forward ($E_+$) and backward propagating fields ($E_-$) as
	\begin{gather}
	E(z)=E_+ \exp(ikz)+E_- \exp(-ikz)
	\label{Eq: Norm2}
	\end{gather}

	As we substitute Eq. (\ref{Eq: Norm2}) in Eq. (\ref{Eq: Norm1}), we can readily arrive at the coupled-mode equations (CMEs) that characterize the wave propagation in the proposed system provided that a slowly varying envelope approximation is applied during the mathematical manipulations. The final form of the CMEs can be written as \cite{pelinovsky2002stable,pelinovsky2002stableI,komissarova2019pt}
	\begin{gather}\label{Eq: Norm3}
		\begin{aligned}
			\color{blue}+& i\left(\frac{d E_{+}}{d \zeta}\right)+\bar\delta(\zeta)+(\kappa + g) E_-+\gamma_{nl}\left(\left|E_+\right|^{2}+2\left|E_-\right|^{2}\right) E_+ \\
			&+\gamma_{K}\left(\left|E_-\right|^{2}+2\left|E_+\right|^{2}\right) E_-+\gamma_{K} E_+^{2} E_-^{*}=0 \\
			-& i\left(\frac{d E_-}{d \zeta}\right)+\bar\delta(\zeta)+(\kappa - g) E_++\gamma_{nl}\left(\left|E_-\right|^{2}+2\left|E_+\right|^{2}\right) E_- \\
			&+\gamma_{K}\left(\left|E_+\right|^{2}+2\left|E_-\right|^{2}\right) E_++\gamma_{K} E_-^{2} E_+^{*}=0
		\end{aligned}
	\end{gather}

	The mathematical notations used in Eq. (\ref{Eq: Norm3}) are as follows: $\zeta$-normalized propagation coordinate, $\kappa$-coupling coefficient which describes the amount of power coupled between the transmitted and reflected signals, $g$-gain and loss term which accounts for the effect of $\mathcal{PT}$-symmetry, $\gamma_{nl}$-cubic nonlinearity, $\gamma_{K}$- modulation of the Kerr nonlinearity which is averaged over the layered structure, and $\bar\delta$- modified detuning parameter which accounts for the effect of chirping as well as detuning. The mathematical forms of these parameters are described below,
	\begin{gather}
\nonumber	\bar\delta(z)= \delta+C(\zeta-L/2)/L^2, \quad \delta=2 \pi n_0 \left(\cfrac{1}{\lambda_0}-\cfrac{1}{\lambda_b}\right) \\\nonumber \lambda_b = 2 n_0 \Lambda, \quad \kappa=\pi n_{1R}/\lambda_0, \quad g=\pi n_{1I}/\lambda_0, \\ \gamma_{nl} = 2 \pi n_{nl}/\lambda_0 \quad \text{and} \quad \gamma_{K} = 2 \pi n_{K}/\lambda_0
	\end{gather}
Here $C$, $\delta$, $\lambda_b$, and $\lambda_0$ indicate the normalized chirping, detuning parameter, Bragg and operating wavelengths, respectively. Note that Eq. (\ref{Eq: Norm3}) holds good when the light incidence direction is left. If the light incidence direction is right, then the coupled mode equations are modified as

\begin{gather}\label{Eq: Norm3.1}
	\begin{aligned}
		+& i\left(\frac{d E_{+}}{d \zeta}\right)+\bar\delta(\zeta)+(\kappa - g) E_-+\gamma_{nl}\left(\left|E_+\right|^{2}+2\left|E_-\right|^{2}\right) E_+ \\
		&+\gamma_{K}\left(\left|E_-\right|^{2}+2\left|E_+\right|^{2}\right) E_-+\gamma_{K} E_+^{2} E_-^{*}=0 \\
		-& i\left(\frac{d E_-}{d \zeta}\right)+\bar\delta(\zeta)+(\kappa + g) E_++\gamma_{nl}\left(\left|E_-\right|^{2}+2\left|E_+\right|^{2}\right) E_- \\
		&+\gamma_{K}\left(\left|E_+\right|^{2}+2\left|E_-\right|^{2}\right) E_++\gamma_{K} E_-^{2} E_+^{*}=0
	\end{aligned}
\end{gather}

It is assumed that the normalized length and coupling coefficient are fixed (throughout the paper) as $L = 2$ and $\kappa=4$ (unless specified). Similarly, the value of gain and loss coefficient is kept at $g=3.75, 5, 4$ for the unbroken, broken $\mathcal{PT}$-symmetric regimes, and unitary transmission point, respectively. The transmitted field emanating at the other end of the fiber will have a measurable intensity and it is taken to be $P_1(L)$. Then, the OB curves that describe the nonlinear input-output characteristics of the system can be obtained by plotting the input intensity ($P_0$) against the transmitted intensity($P_1(L)$) under the variation of different control parameters. Implicit Runge-Kutta fourth-order method is one of the efficient numerical schemes used by many researchers working with nonlinear FBGs and so we adhere to the same.  The boundary conditions which we have chosen are
	\begin{gather}
	E_+(0)=E_0, \text{and} \quad E_-(L)=0
	\end{gather}

	\section{Light launching direction: Left}\label{Sec:III}

Before we present our results, we would like to recap some of the basic concepts related to the optical bistability phenomenon. Formally, it is defined as a nonlinear physical process in which two output states are generated for a given input state provided that a sufficient feedback is given to the system. In the absence of any one of these requirements (nonlinearity and feedback), OB will not occur. As we tune the input intensity of the laser, the system will behave linearly at first till a threshold power value, say $P_{up}^{th}$, is reached as shown in Fig. \ref{fig0B}. The corresponding linear curve between $0<P_0<P_{up}^{th}$ represents the first stable state (AB). Once the intensity crosses this threshold value, then it will undergo a sudden jump (to C) in the magnitude of the output intensity $P_1(L)$. This action is generally referred to as the up-switching phenomenon. Post the occurrence of this switching action, it remains to behave linearly (again) to any variations in the input power and this is taken to be the second stable state (CD). Once again the switching can occur in the perspective of highly nonlinear glasses like chalcogenide and hence it is named as optical multistability phenomenon which is not the subject of interest in this article. If one intends to return to the original state (second stable state to first), then the input intensity can be slowly reduced. But due to the inherent property of the device, the down-switching (jump from second to the first stable state) will occur at a different power level $P_{down}^{th}$ far away from point B, say E. Between these two stable states, an unstable state (BE) will appear in the numerical output.
	\subsection{Unbroken \texorpdfstring{$\mathcal{PT}$}--symmetric regime}
	\label{Sec:3a}
	It should be recalled that when the magnitude of the coupling coefficient ($\kappa$) dominates the magnitude of the gain and loss parameter ($g$), then the system is said to be operating in the unbroken $\mathcal{PT}$-symmetric regime.
	\subsubsection{Type of Nonlinearity: Self-focusing ($\gamma_{nl}, \gamma_K > 0$)}
\begin{figure}[t]
	\centering	\includegraphics[width=0.5\linewidth]{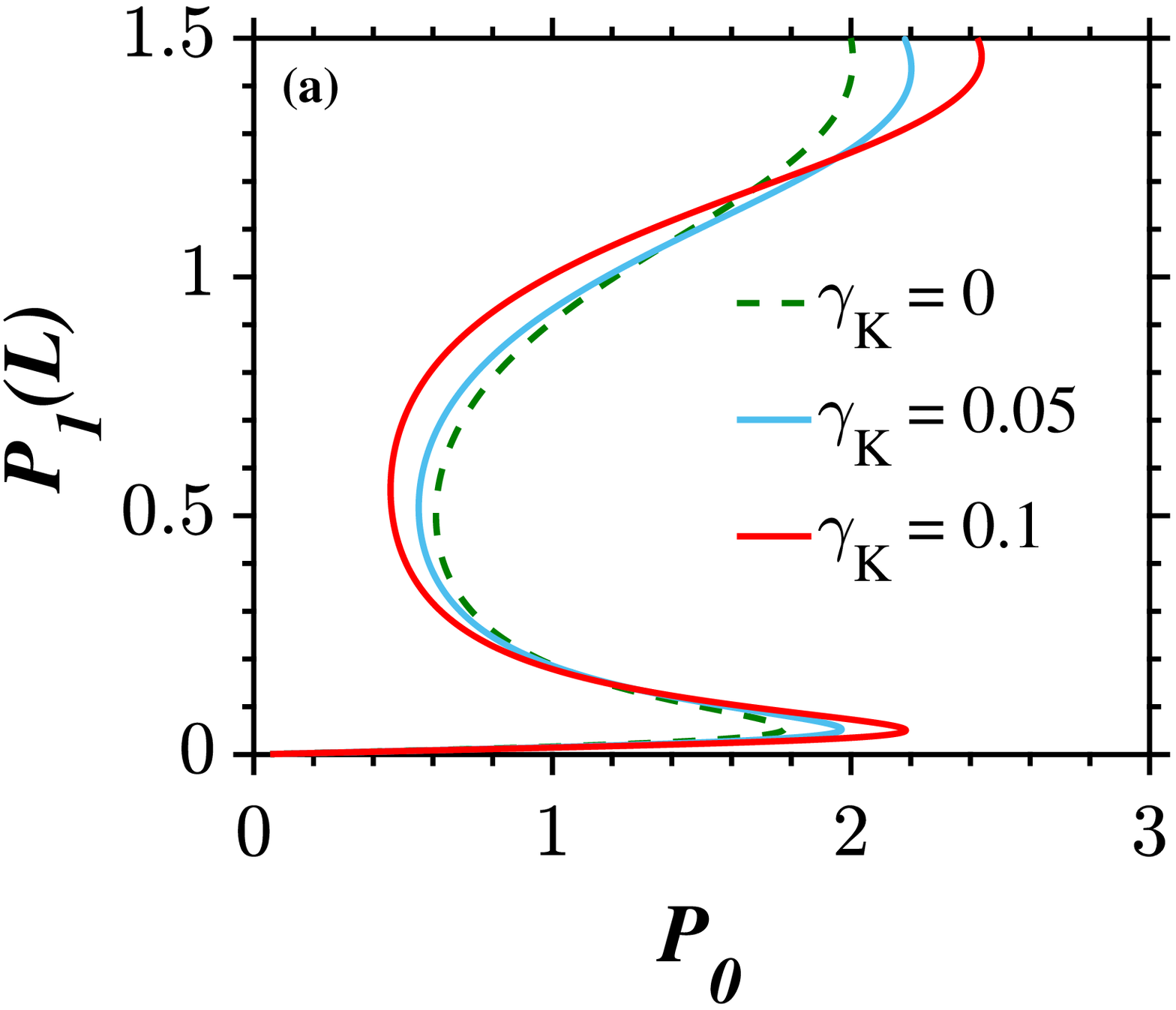}\includegraphics[width=0.5\linewidth]{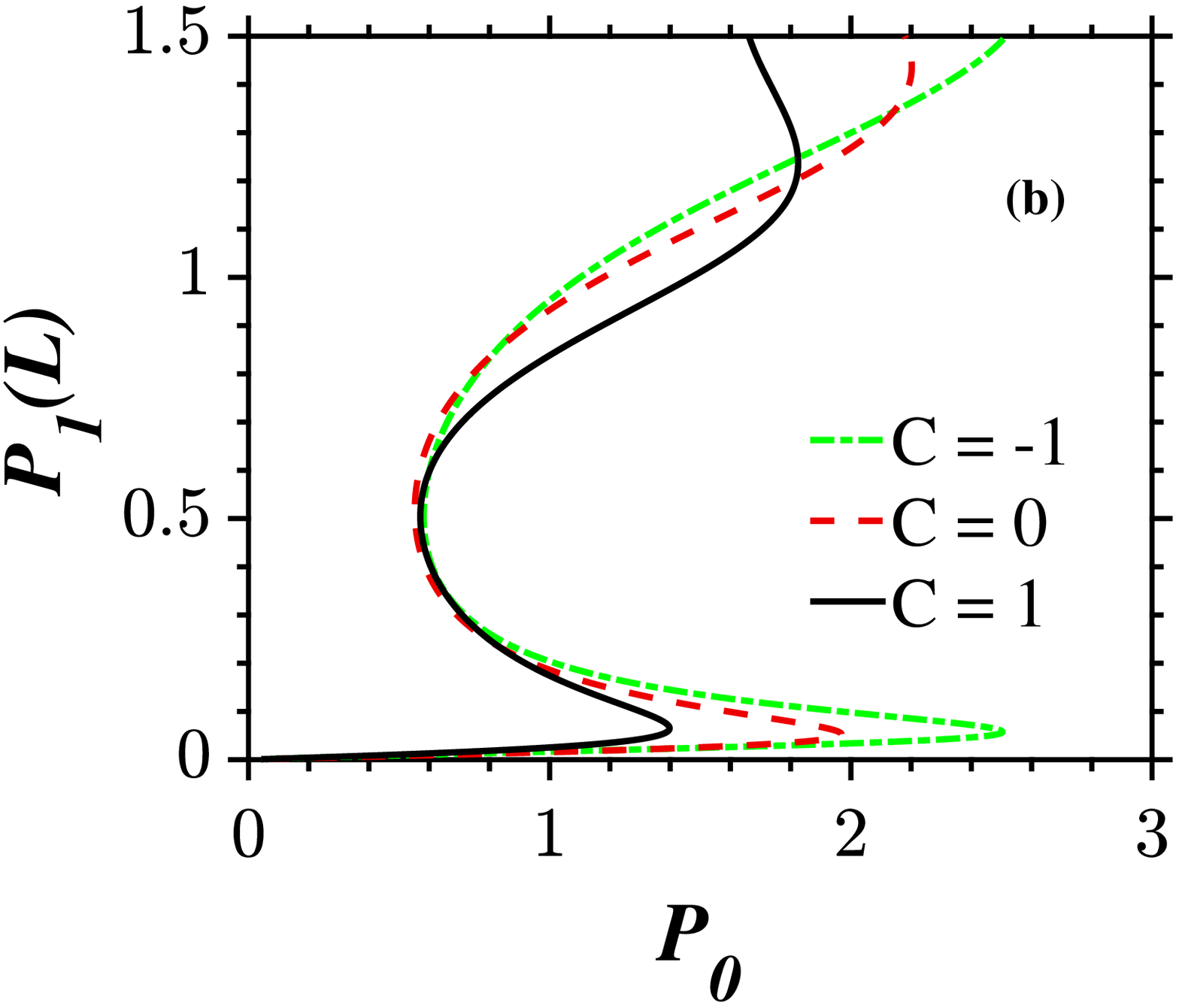}\\\includegraphics[width=0.5\linewidth]{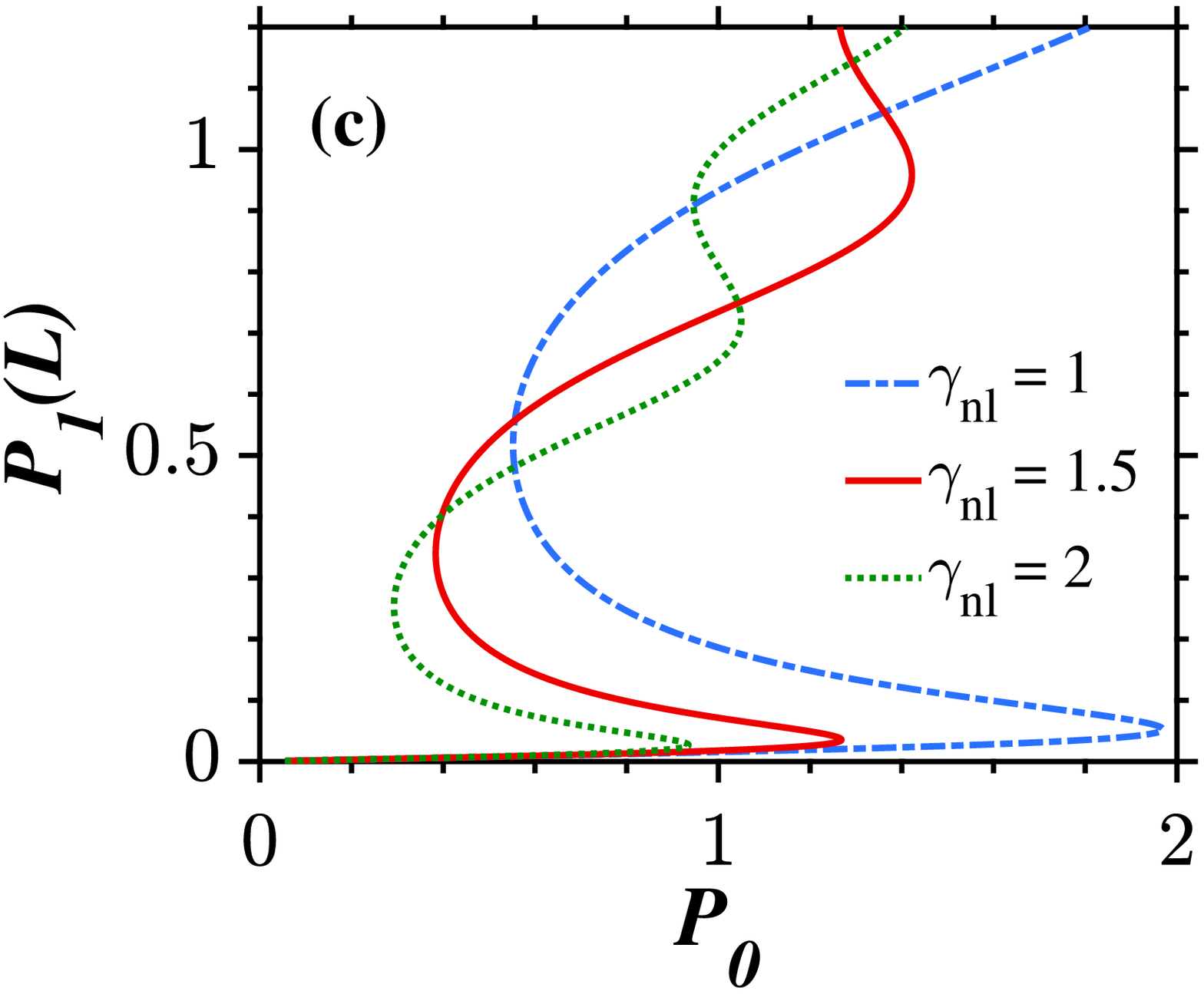}
	\caption{OB in a CPTFBG operating in the unbroken $\mathcal{PT}$-symmetric regime with modulation of the Kerr nonlinearity ($\gamma_K$) and self-focusing cubic nonlinearity ($\gamma = 1$)  for the left light incidence. (a) Role of $\gamma_K$ at $\delta = C = 0$. (b) Effect of variations in the chirping parameter ($C$) at $\delta=0$ and $\gamma_K=0.05$. The role of self-focusing nonlinearity ($\gamma_{nl}>0$) on the switching intensities is depicted in (c). }
	\label{fig1}
\end{figure}
Since different combinations of signs among the chirping, detuning, nonlinear parameter, and nonlinear modulation term are feasible, it is important to find the regime under which the system gives rise to optical steering at low powers. To do so, we fix the value of the nonlinearity parameter to be $\gamma_{nl} = 1$ in Fig. \ref{fig1} which means that the nonlinearity is of self-focusing type. From Fig. \ref{fig1}(a), one can infer that any increase in the value of the modulation of Kerr nonlinearity parameter ($\gamma_K$) will increase the power required ($P_{up}^{th}$ and $P_{down}^{th}$) to switch between the stable states and the width of hysteresis starts to grow with an increase in the value of $\gamma_K$ as shown in Fig. \ref{fig1}(a). This clearly explains that the combination of self-focusing nonlinearity along with self-focusing nonlinear modulation term is not the preferable regime for realizing all-optical switches. In our previous work, we have already demonstrated that any sign mismatch between the type of nonlinearity and chirping parameter will result in switching at higher powers \cite{raja2019nonlinear}. This conclusion holds good in the presence of modulation of Kerr nonlinearity term ($\gamma_K$) also as shown in Fig. \ref{fig1}(b). In other words, the combination of self-focusing nonlinearity, modulation of nonlinear parameters, and positive chirping can reduce the switching power considerably, which will be proven in the next subsection. The role of nonlinearity parameter on the switching is depicted in Fig. \ref{fig1}(c) and we infer that with increase in the value of $\gamma_{nl}$ the intensity required to switch between the states decreases. We would like to note that the value of the length of the device ($L$) and the coupling coefficient ($\kappa$) should be chosen carefully so that the feedback to the system is sufficient without which the optical bistability phenomenon cannot occur \cite{raja2019multifaceted}. Through a series of numerical experiments we found that if the value of $L$ is too large ($L > 5$) and the coupling coefficient ($\kappa$) is comparatively less than $L$ [$L>>k$], the system produces undesirable bistable states which indicate the onset of instability. The instability can be managed by suitably increasing the value of the nonlinear coefficient ($\gamma_{nl}$) and keeping the value of gain and loss parameter ($g$) closer to $k$. This once again proves that tuning the value of the nonlinearity and the value of gain and loss can essentially suppress the instability  \cite{huang2014pt, luz2019robust}.
	\subsubsection{Type of Nonlinearity: Self-defocusing ($\gamma_{nl}, \gamma_K < 0$)}
\begin{figure}[t]
	\centering	\includegraphics[width=0.5\linewidth]{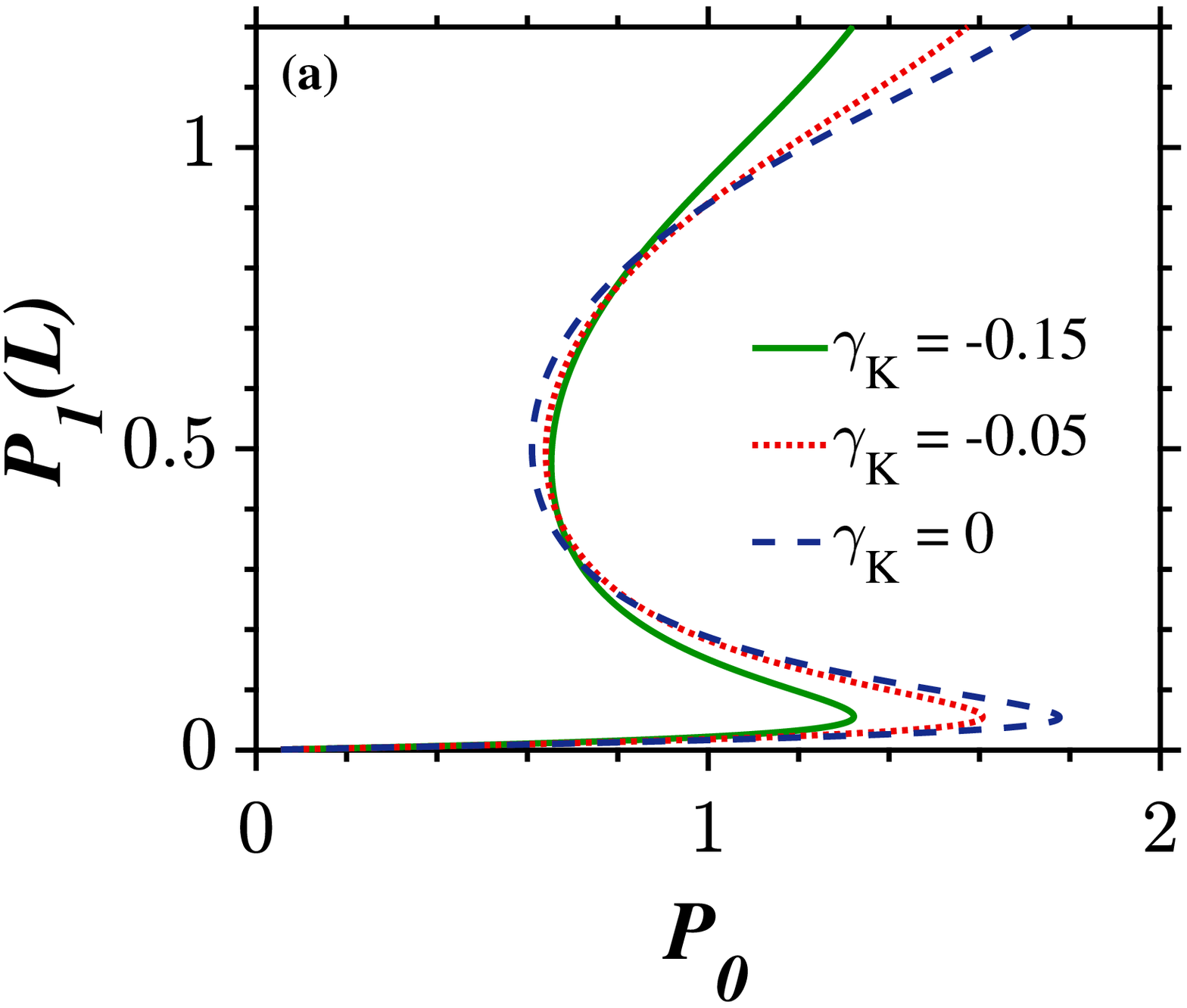}\includegraphics[width=0.5\linewidth]{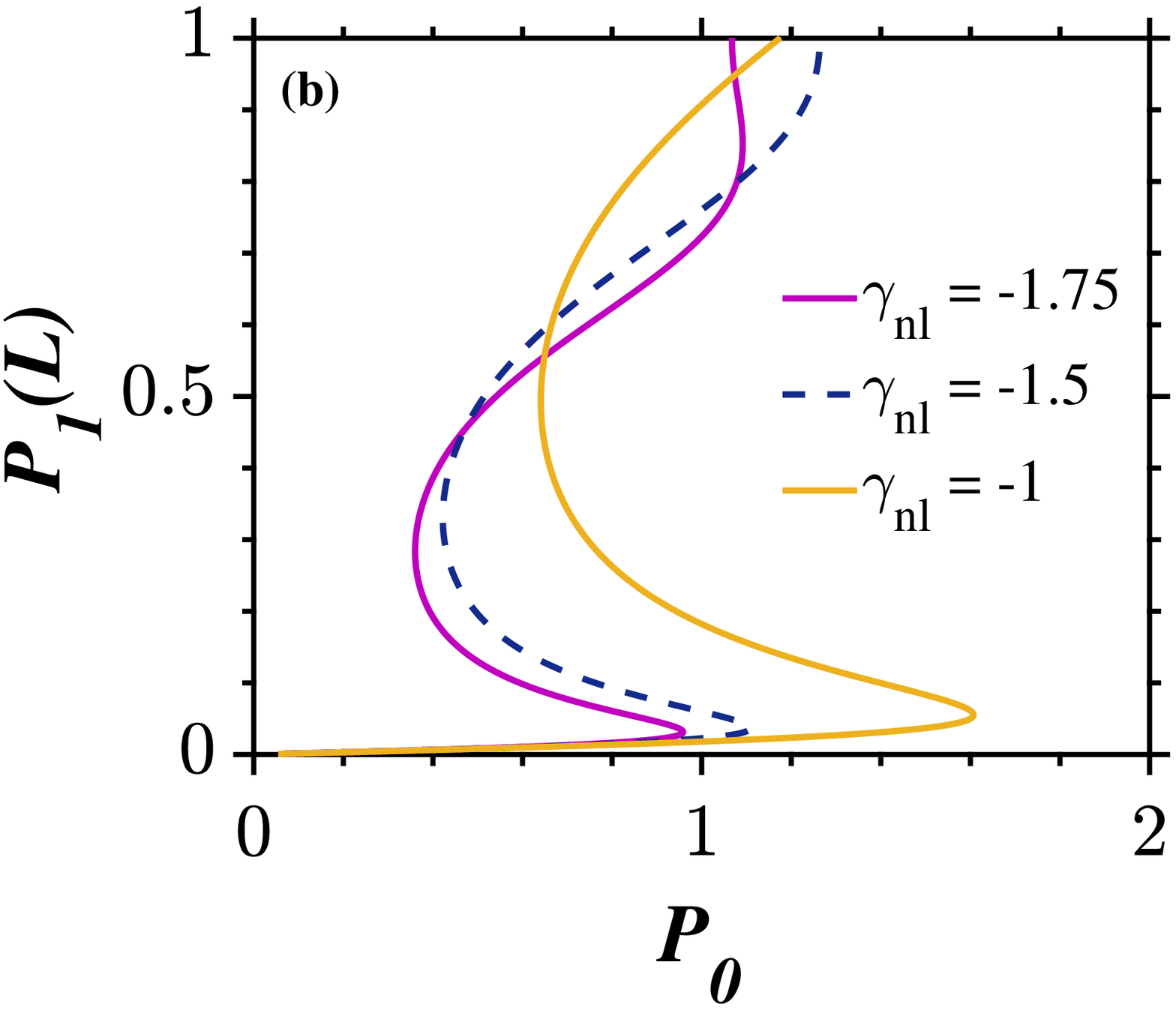}\\\includegraphics[width=0.5\linewidth]{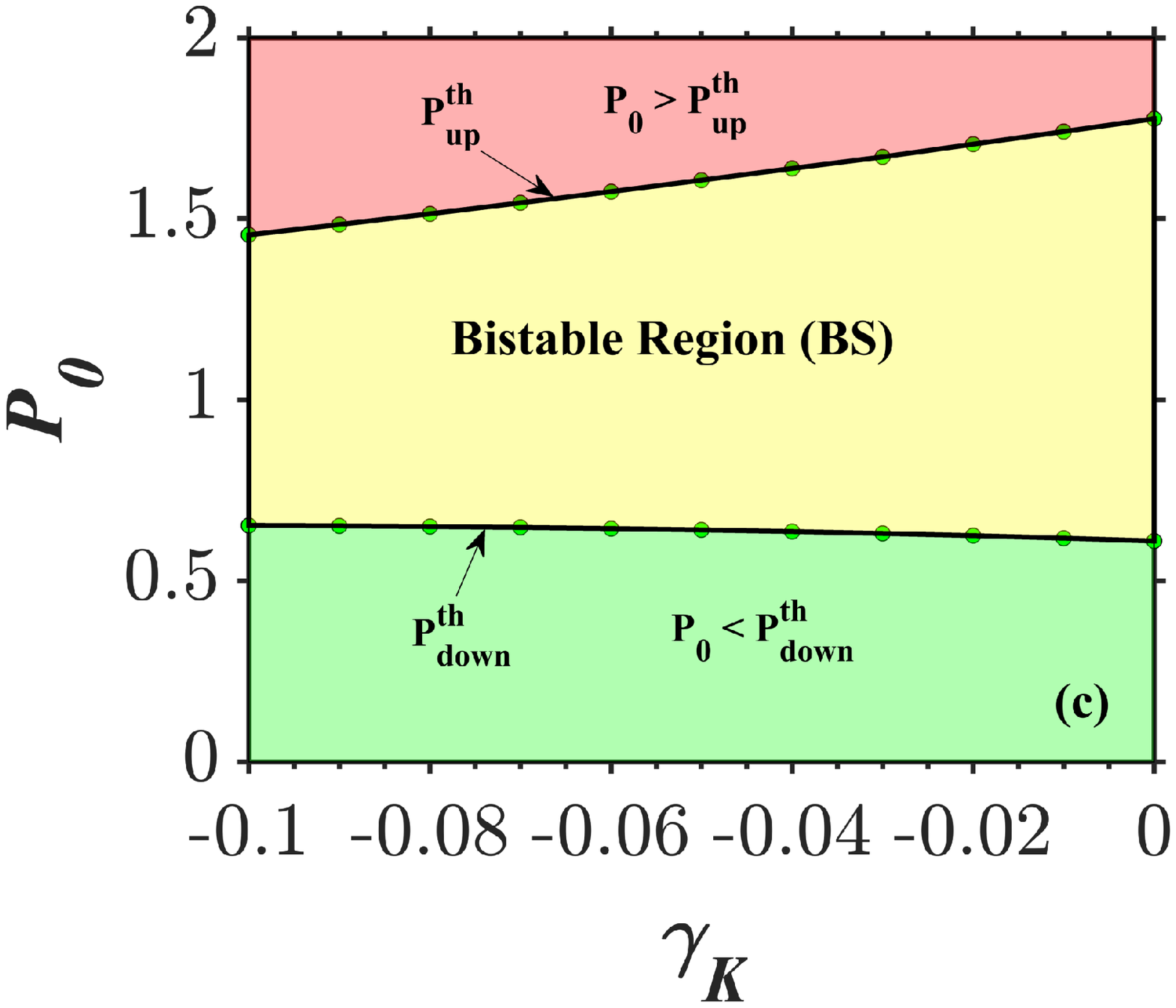}\includegraphics[width=0.5\linewidth]{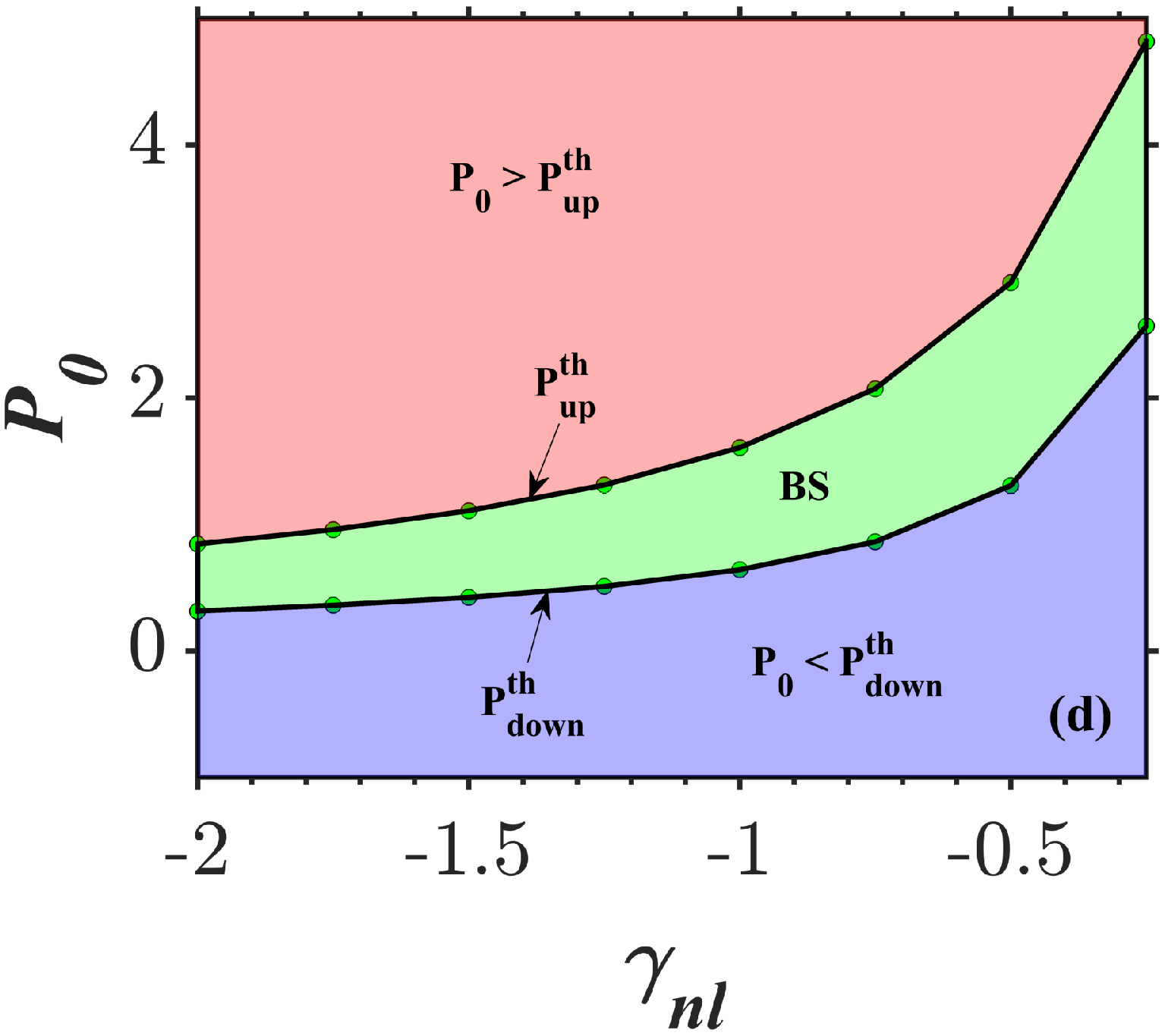}
	\caption{Plots illustrating the optical bistability dynamics under self-defocusing nonliear regime ($\gamma = -1$) at $\delta= C = 0$. (a) Role of modulation of the Kerr nonlinearity parameter ($\gamma_K$) on switching. (b) Effect of variations in the cubic nonlinearity ($\gamma_{nl}$)  at $\gamma_K = -0.05$ on switching. Illustration of variation in the switching intensities with respect to (c) the modulation of Kerr nonlinearity ($\gamma_K$) and (d) nonlinearity ($\gamma_{nl}$) parameters. }
	\label{fig2}
\end{figure}
We have concluded from Fig. \ref{fig1}(a) that operating the system with the self-focusing nonlinear modulation parameter is not an ideal choice to reduce the switching powers. For this reason, we look into the role of $\gamma_K$ under the assumption that it is of a self-defocusing type ($\gamma_{K}=-1$). As we tune the value of $\gamma_K$, we can observe that the power required to switch between the stable states decreases as shown in Fig. \ref{fig2}(a). Comparing Figs. \ref{fig1}(a) and \ref{fig2}(a), we  arrive at a conclusion that the inclusion of modulation of Kerr nonlinearity (FWM) term ($\gamma_K$) is beneficial only if it is assumed to be of self-defocusing type. The thumb rule to realize an all-optical switch with low switching powers is to use a material with a high value of nonlinearity in magnitude and not in sign ($\gamma_{nl}$)  \cite{raja2019multifaceted}  and this is confirmed from Fig. \ref{fig2}(b). The negative sign before the nonlinearity parameter indicates that the type of nonlinearity is self-defocusing. As we decrease the value of the modulation of the Kerr nonlinearity ($\gamma_K$) parameter, the value of $P_{up}^{th}$ reduces gradually as seen in Fig. \ref{fig2}(c). One can infer from Fig. \ref{fig2}(d) that the bistable region gets narrowed as the value of the nonlinear coefficient is increased (in magnitude) which confirms the fact that more negative value of the nonlinear coefficient, lower will be the intensity required to switch. 

\subsubsection{Role of chirping ($C$) and detuning ($\delta$) parameters for  $\gamma_{nl}, \gamma_K < 0$}
\begin{figure}[t]
	\centering	\includegraphics[width=0.5\linewidth]{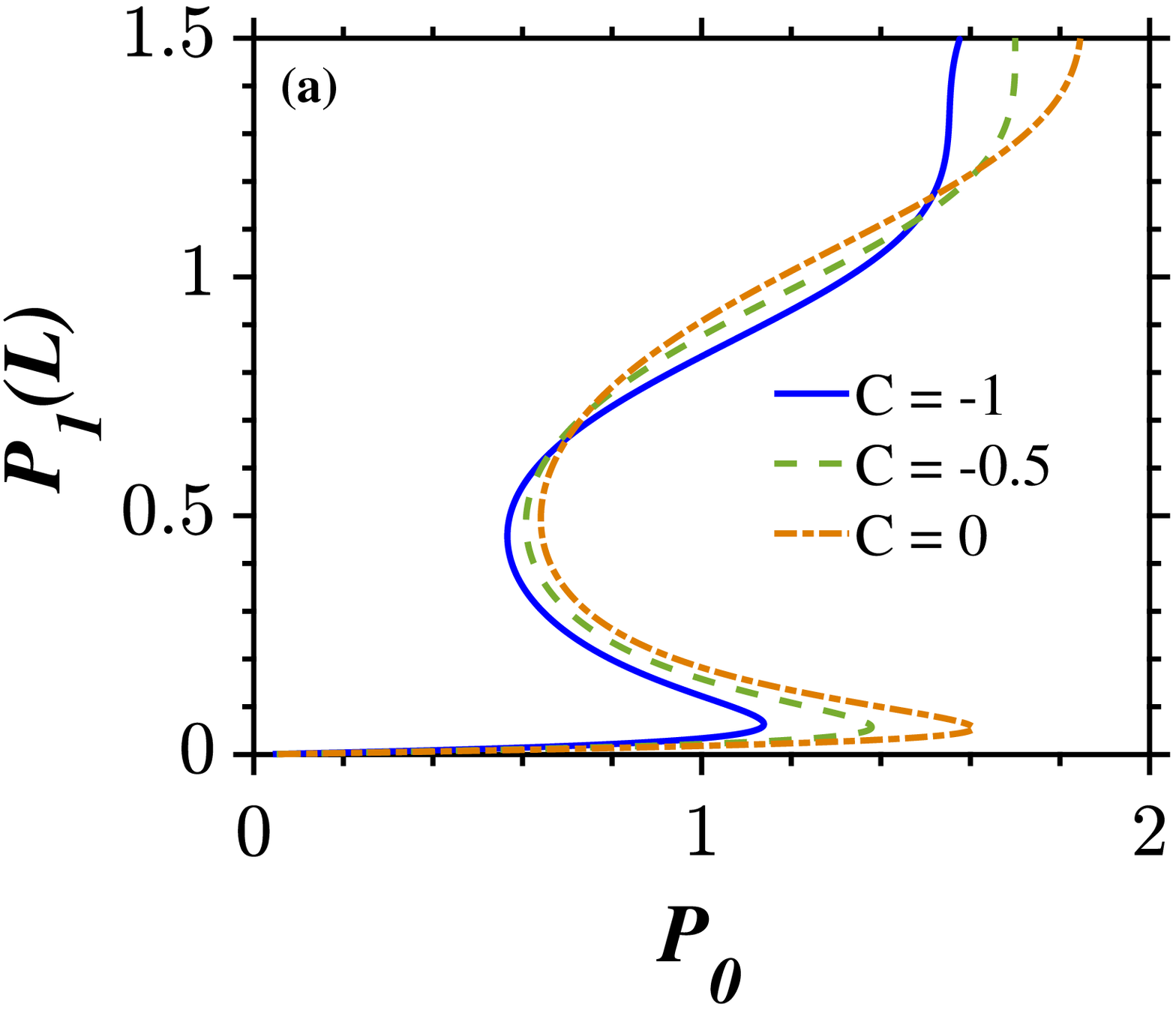}\includegraphics[width=0.5\linewidth]{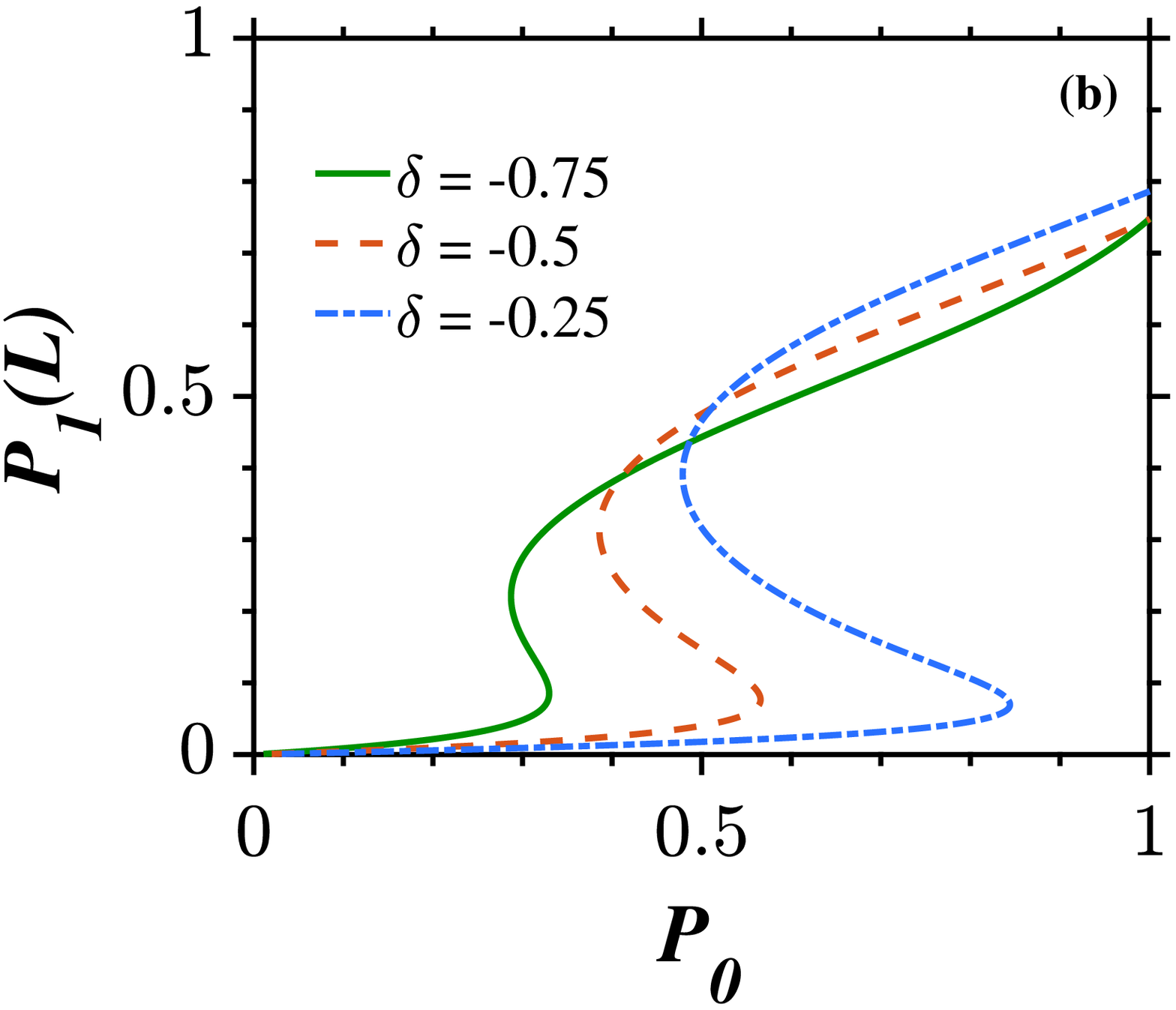}\\\includegraphics[width=0.5\linewidth]{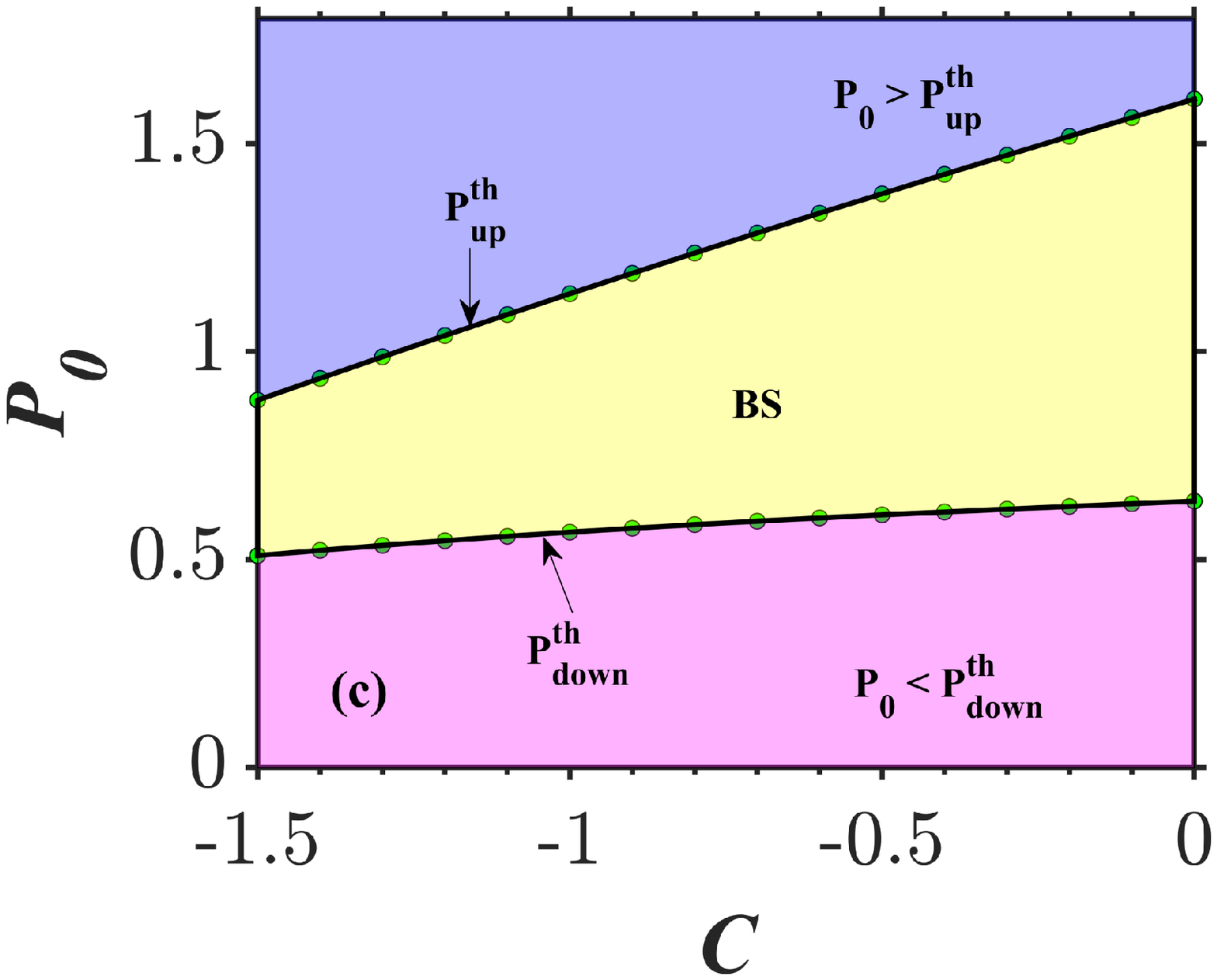}\includegraphics[width=0.5\linewidth]{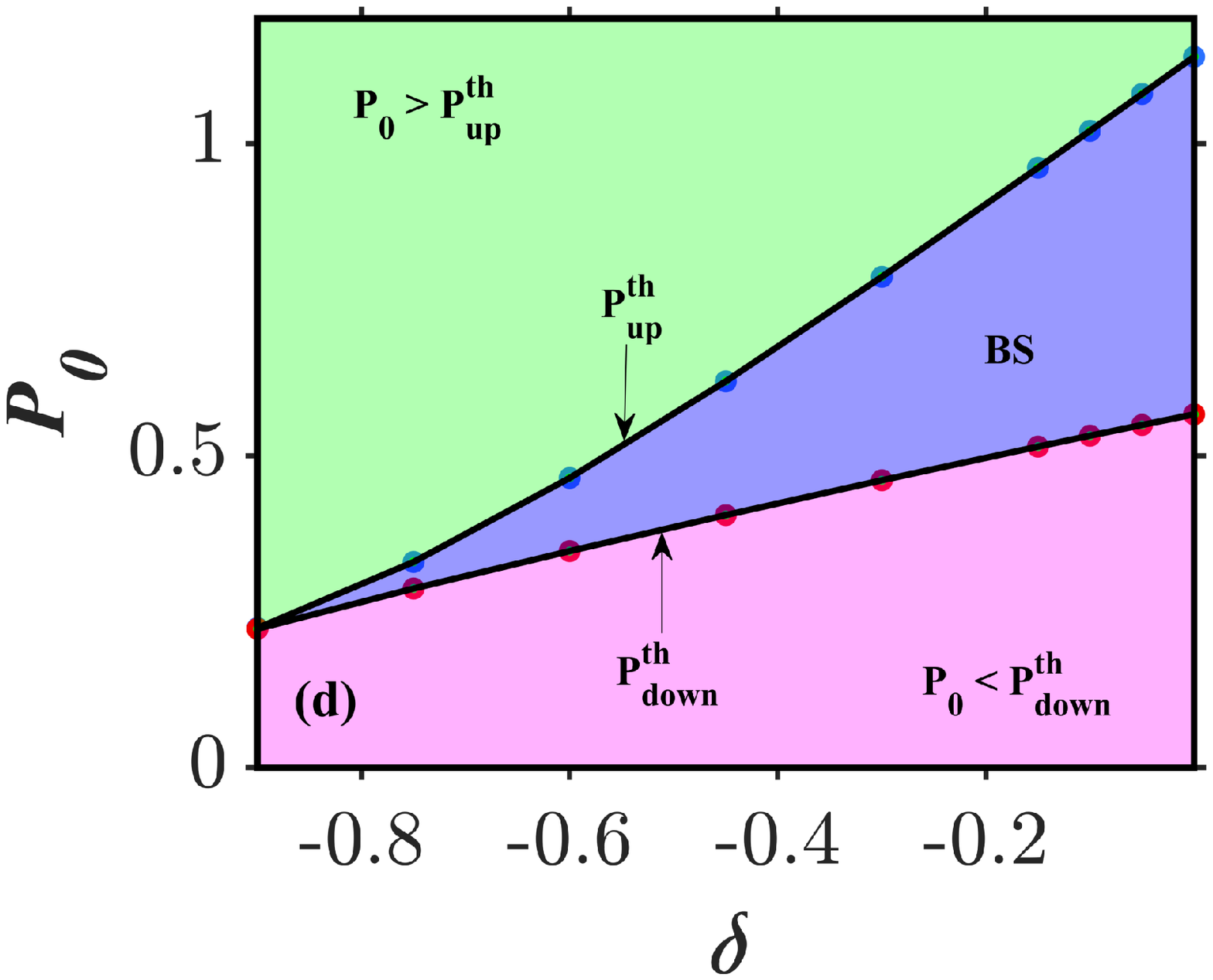}
	\caption{OB in a CPTFBG operating in the unbroken $\mathcal{PT}$-symmetric regime under left light incidence direction in the presence of self-defocusing cubic nonlinearity ($\gamma_{nl} = -1$) and modulation of Kerr nonlinearity ($\gamma_K = -0.05$). (a) Role of negative chirping parameter ($C < 0$) at $\delta = 0$. (b) Effect of varying the detuning parameter ($\delta$) on the OB curve at $C = -1$. (c) and (d) portray the effects of continuous variation of the chirping parameter ($C$) and detuning ($\delta$), respectively, on the switching intensities ($P_0$). }
	\label{fig3}
\end{figure}

Hereafter, both the Kerr nonlinearity and FWM terms are assumed to be self-defocusing nonlinear ones (unless specified) throughout the paper, since such a scheme results in low power steering. In this section, the Kerr and modulated Kerr nonlinearity (FWM) coefficients are taken to be $\gamma_{nl} = -1$ and $\gamma_K = -0.05$, for simplicity, and the role of chirping nonuniformity ($C$) is investigated under the influence of the detuning parameter. Figure \ref{fig3}(a) depicts the role of the negative detuning parameter ($C<0$) on the bistable switching. Compared to the operation in the absence of chirping (indicated by dashed-dotted lines), the inclusion of the negative detuning parameter favors bistable switching at very low powers as inferred from Fig. \ref{fig3}(a). Since all the laser sources exhibit a fine lasing linewidth, the discrepancy between the Bragg ($\lambda_b$) and operating($\lambda_0$) wavelengths must have a definite impact on the nonlinear dynamics exhibited by the system, and the parameter which describes the difference between the two wavelengths is designated as the detuning parameter ($\delta$). The system is said to be negatively detuned if the operating wavelength lies above the Bragg wavelength ($\lambda_0 > \lambda_b$) and vice-versa. In the presence of a large value of negative detuning parameter ($\delta < 0$), the system supports low power all-optical switching as depicted in Fig. \ref{fig3}(b). But it should be remembered that detuning far away from Bragg wavelength or too much of reduced frequency of the grating [as a result of the high value of negative detuning parameter ($C<0$)]  will also reduce the effective feedback supplied to the system and so the bistable states cannot occur. Before moving further, we would like to introduce an important feature exhibited by chirped FBG structure, namely the spectral span. This parameter refers to the range of operating wavelengths over which we can observe a typical bistable phenomenon in the input-output characteristics curve. Since chirping refers to the variation in the spatial frequency of the grating along the propagation direction, any variation in the chirping parameter could play a pivotal role in altering the range of wavelengths over which OB occurs as depicted in Fig. \ref{fig3}(c). Alternatively, one can tune the spectral range to the desired regime by judiciously manipulating the detuning parameter as shown in Fig. \ref{fig3}(d).  

\subsection{Broken \texorpdfstring{$\mathcal{PT}$}--symmetric regime}\label{Sec:3b}
\begin{figure}[t]
	\centering	\includegraphics[width=0.5\linewidth]{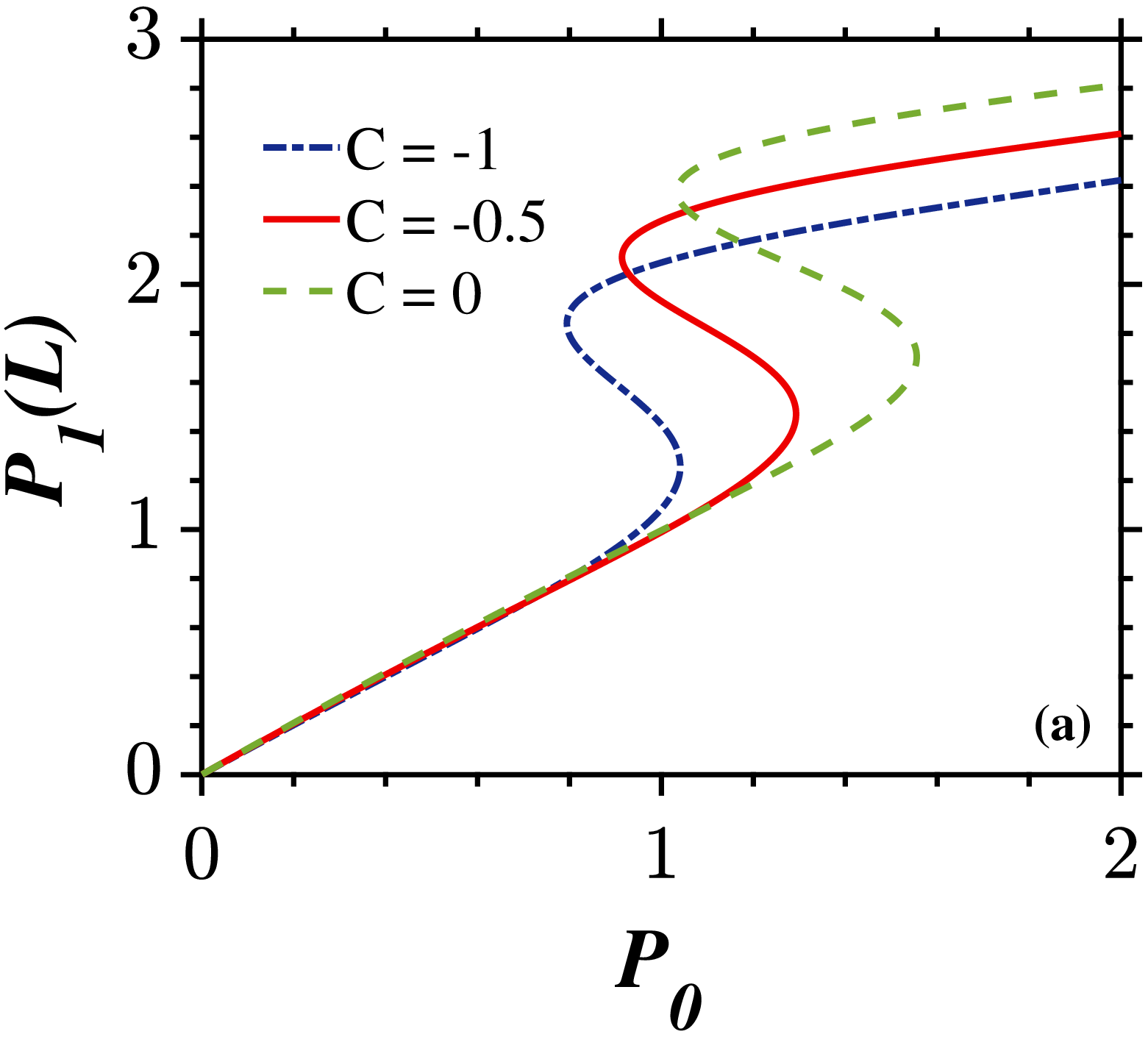}\includegraphics[width=0.5\linewidth]{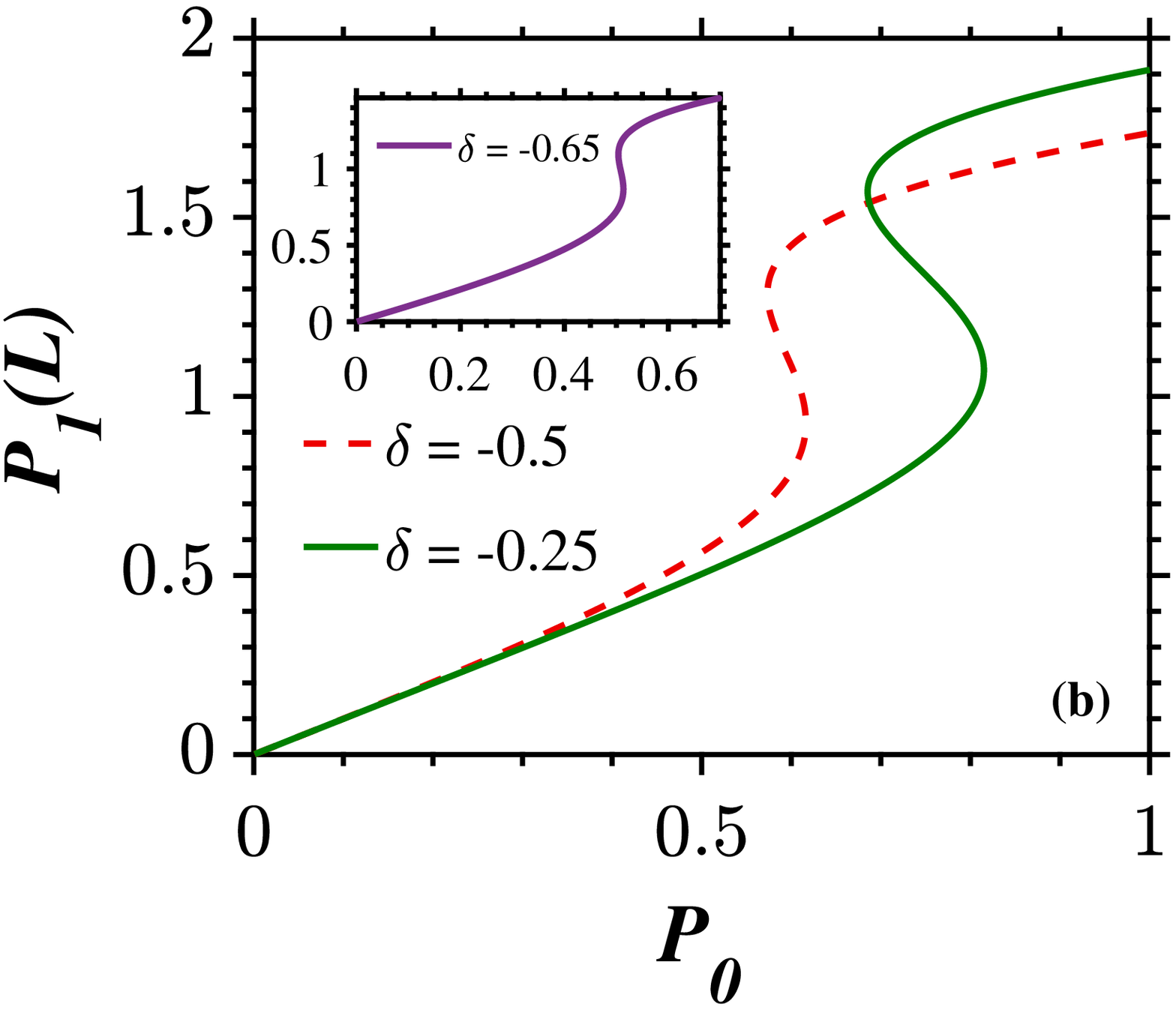}\\\includegraphics[width=0.5\linewidth]{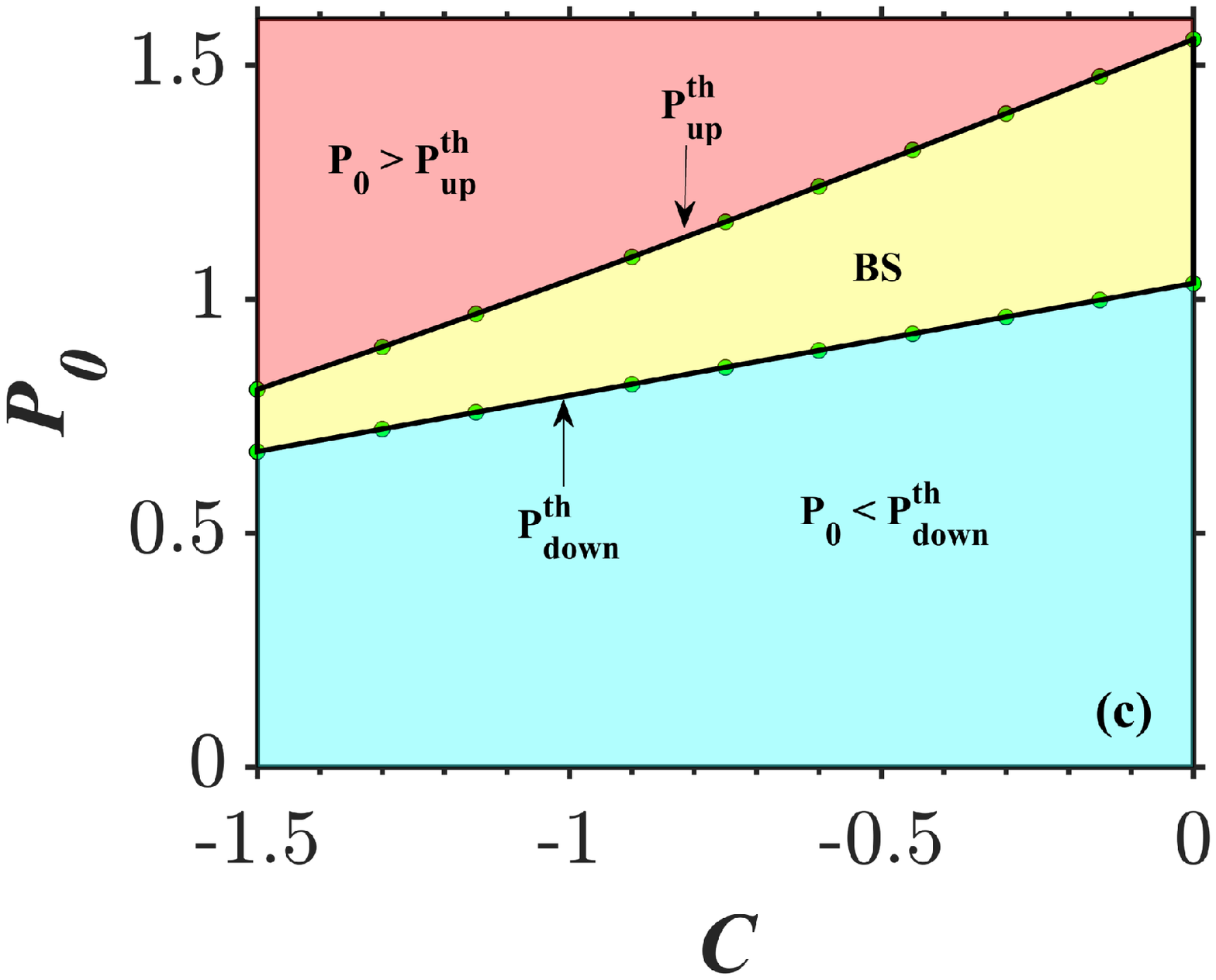}\includegraphics[width=0.5\linewidth]{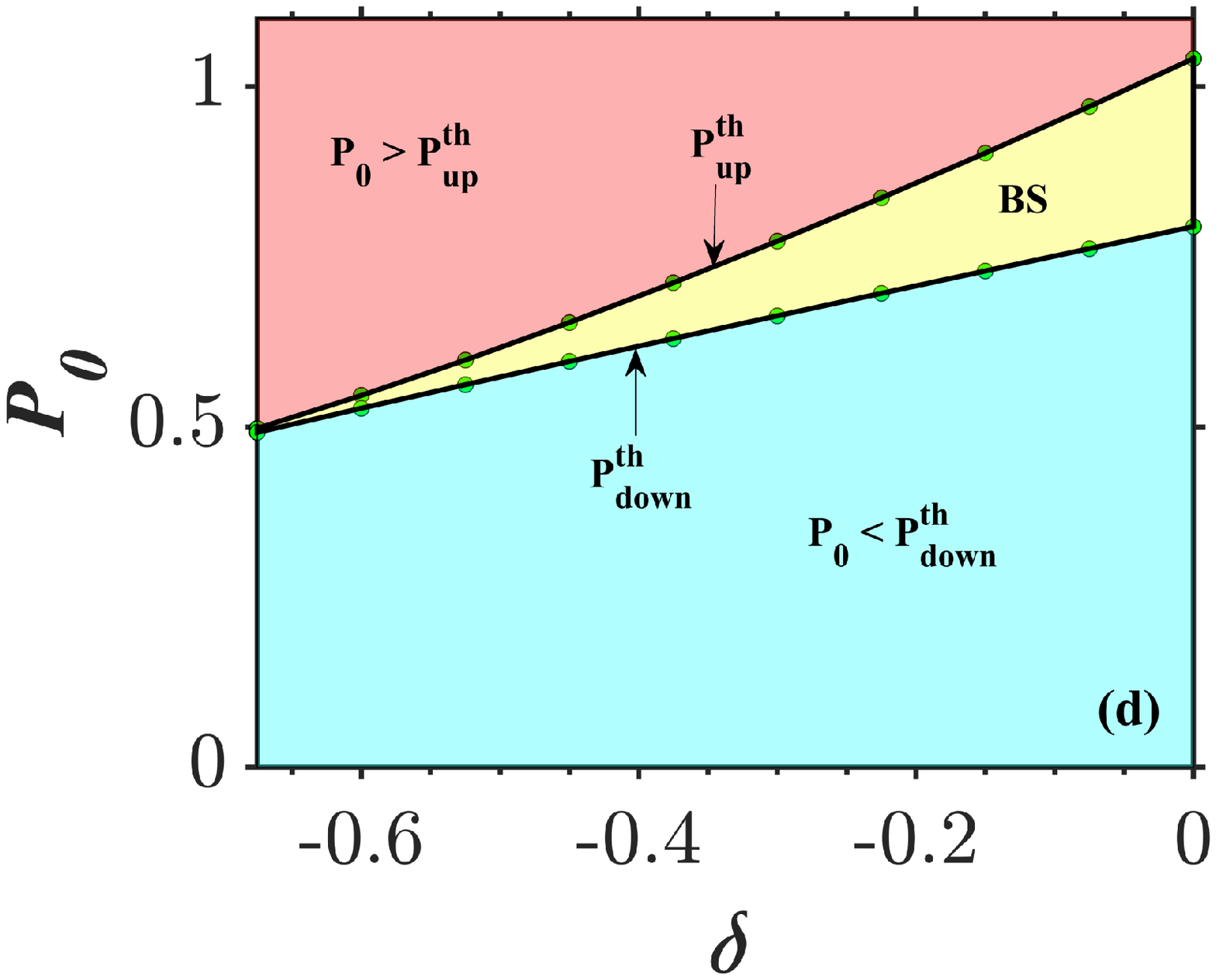}
	\caption{Plots showing the same dynamics with the same system parameters as in Fig. \ref{fig3} except that it is operated in the broken $\mathcal{PT}$-symmetric regime.}
	\label{fig4}
\end{figure}
It is evident from Sec. \ref{Sec:III}\ref{Sec:3a} that under favorable conditions and in the presence of sufficient feedback [the product of coupling coefficient ($\kappa$) and length of the grating ($L$)], the system will support low-power steering provided that the signs of nonlinearity ($\gamma_nl$), detuning ($\delta$), chirping ($C$), and modulation of the Kerr nonlinearity ($\gamma_K$) terms are negative in the unbroken $\mathcal{PT}$-symmetric regime. Mathematically, the condition $g > \kappa$ indicates that the PTFBG is operated in the broken $\mathcal{PT}$-symmetric regime. The high value of nonlinearity alongside the gain supplied to the system may lead to instability under some specific values of system parameters \cite{komissarova2019pt}. But, if the excitation source is a high power continuous wave laser and the system parameters are kept at optimal values, it is possible to obtain stable states in the broken $\mathcal{PT}$-symmetric regime of the nonlinear uniform \cite{raja2019multifaceted} and nonuniform \cite{raja2019nonlinear} PTFBGs. This proves that a judicious selection of the operating parameters is an inevitable component to avoid numerical instabilities.  Exotic light dynamics like the existence of ramp-like stable states were discovered during our studies in the broken $\mathcal{PT}$-symmetric regime \cite{raja2019multifaceted,raja2019nonlinear}. The question that arises here is that whether the ramp-like stable states can continue to evolve in the company of perturbations in the form of modulation of the Kerr nonlinearity ($\gamma_K$) and chirping ($C$). To answer the query, we carry out a numerical experiment with $g = 5$ as shown in Fig. \ref{fig4}. In Fig. \ref{fig4}(a), a sharp increase in the output intensity with respect to the variations in the input intensity is visible in the plots. Thus, the bistable curves corresponding to several chirping values are well isolated from each other by a predictable amount of input intensity. In addition to this feature, one can also notice that there exists a larger overlap of the first bistable state and a high magnitude of horizontal separation between the second stable states corresponding to different spectral components as inferred from Fig. \ref{fig4}(b). It is already well known that these two functionalities offered by the inhomogeneous grating is designated as spectral uniformity parameter.  The presence of these two significant features indicates that the spectral uniformity offered by the proposed system is large enough to operate as optical memory \cite{maywar1998effect,maywar1997transfer,maywar1998low}.The intensity at which switching happens can be used as the control signal to switch between the binary states (On and Off) of a flip-flop \cite{raja2019nonlinear}. What is more surprising is that the degree of spectral uniformity is appreciable in the presence of modulation in the nonlinear index of the grating. From the continuous variation of the curves plotted in Figs.\ref{fig4}(c) and \ref{fig4}(d), we infer that there exists a large range of input powers common to different spectral components before the steering action takes place. 
The power separation between the different components tends to vary according to the magnitude of chirping and detuning parameters. From these two inferences, we can justify the fact that the broken $\mathcal{PT}$-symmetry offers large spectral uniformity \cite{raja2019nonlinear} and the degree of spectral uniformity can be uniquely controlled by varying either the chirping or detuning or both.

\section{Light launching direction: Right}\label{Sec:IV}

Komissarova \emph{et al.} explored the consequences of operating the nonlinear PTFBG under right light incidence for the first time \cite{komissarova2019pt}. Following this work, we discovered that it is possible to realize ultra low power all-optical switches by reversing the direction of light incidence in a nonlinear uniform and nonuniform PTFBG \cite{raja2019multifaceted,raja2019nonlinear}. 	We are interested in finding a novel solution to reduce the switching intensity further. For this reason, we carry out simulations with the same device parameters as used in Fig. \ref{fig3} except that the optical field is now launched into the PTFBG from the rear end.
	\subsection{Unbroken \texorpdfstring{$\mathcal{PT}$}--symmetric regime ($\kappa>g$)}
	\label{Sec:4a}
\begin{figure}[t]
	\centering	\includegraphics[width=0.5\linewidth]{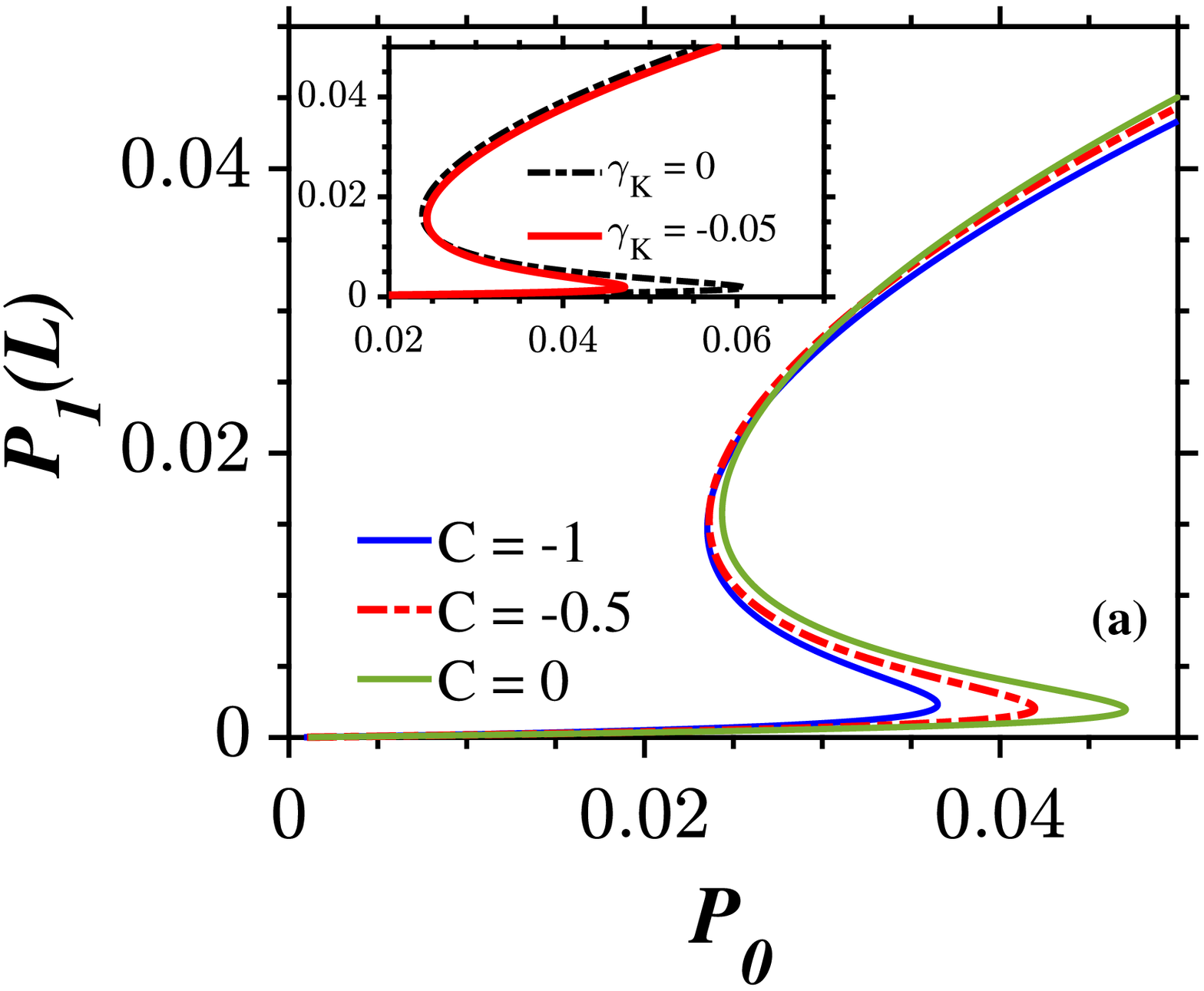}\includegraphics[width=0.5\linewidth]{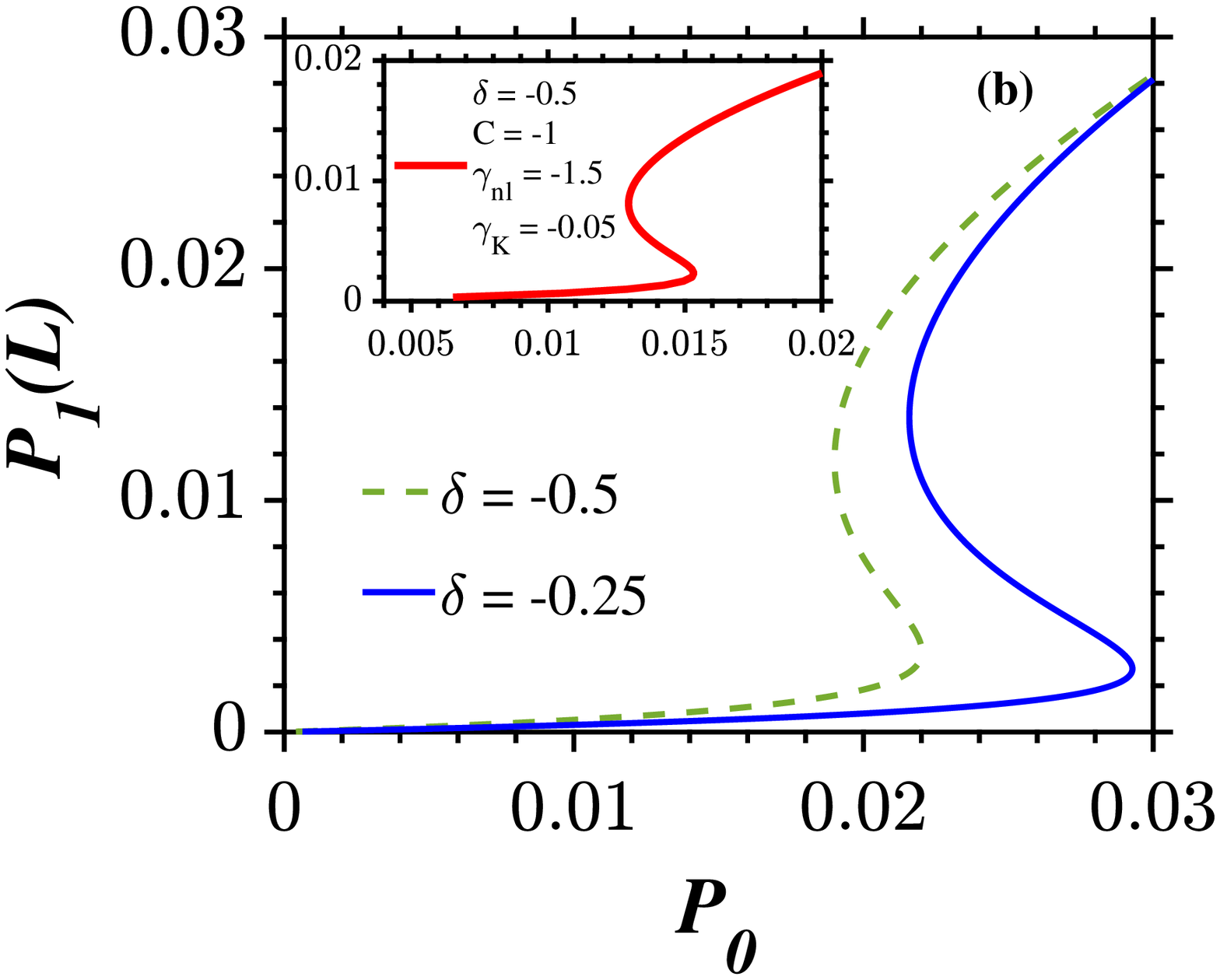}\\\includegraphics[width=0.5\linewidth]{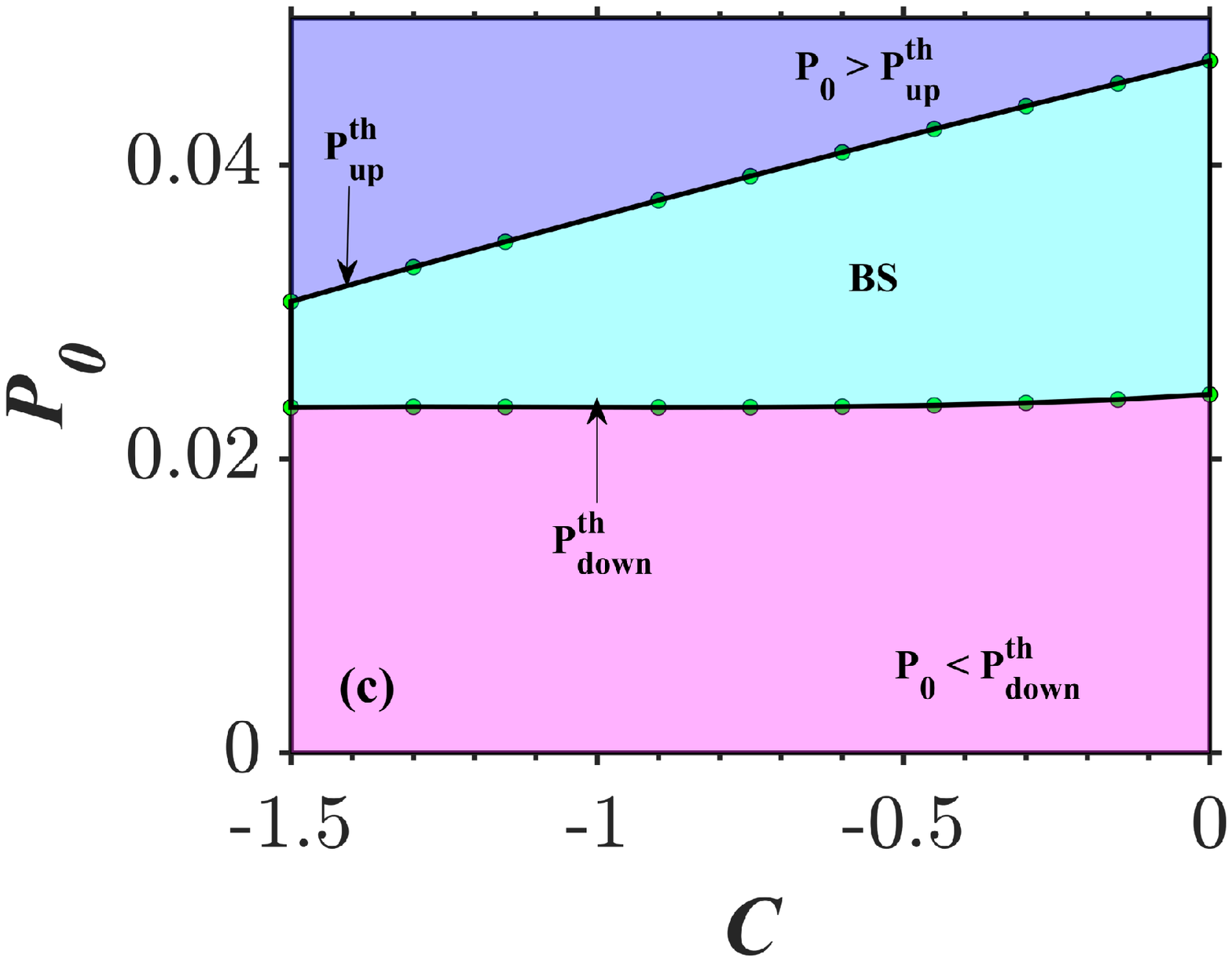}\includegraphics[width=0.5\linewidth]{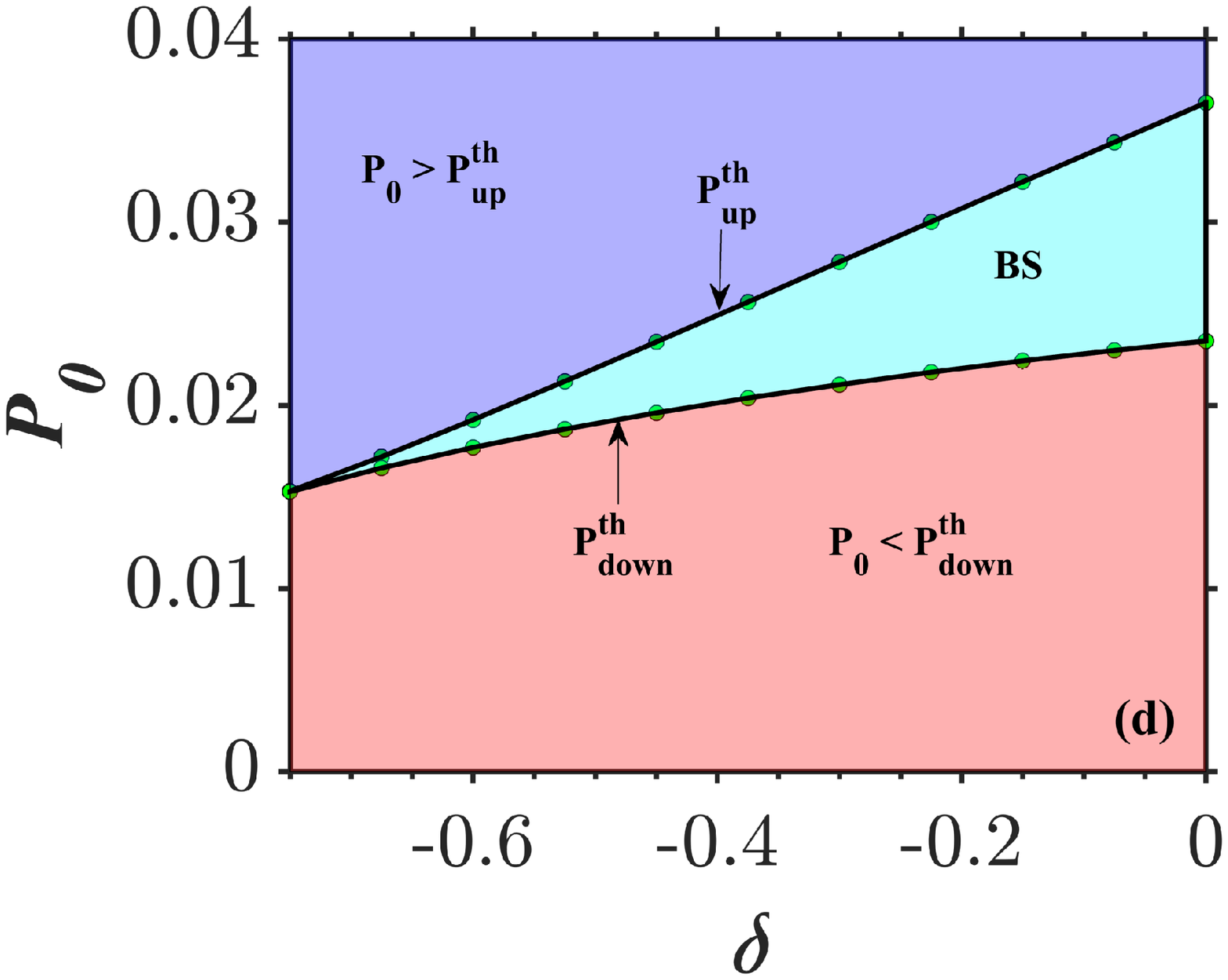}
	\caption{(a) -- (d) Plots showing the same dynamics with the same system parameters as in Fig. \ref{fig3} except that the light launching direction is right. Inset in (a) shows the decrease in the switching intensity in the presence of self-defocusing modulation of Kerr nonlinearity parameter ($\gamma_K$). Inset in (b) depicts the novel switching dynamics occurring at a very low power ($P_0 ~ 0.015$) }
	\label{fig5}
\end{figure}

	The inset in Fig. \ref{fig5}(a) compares the variation in the input intensities required to switch between the two stable states in the absence and presence of the modulation in the nonlinear index profile. Surprisingly, the power required is reduced more than \emph{1.25 times} (roughly) in the presence of self-defocusing modulation of the Kerr nonlinearity term ($\gamma_K = -0.05$). Further, by decreasing the spatial frequency ($C = -1$) along the propagation direction $z$, the switching intensity can be reduced to half of its original value which is supposed to be a dramatic decrement in the intensity required for the steering to occur between the stable states as shown in Fig. \ref{fig5}(a).  The theory of local field enhancement for the right light incidence direction in a grating with gain and loss  should be recalled at this juncture to explain the physical reason behind such a drastic decrease in the input intensities \cite{kulishov2005nonreciprocal, raja2019nonlinear}. In simple words,  the location of field maxima inside the gain regions of the PTFBG is favored in the presence of right light incidence, whereas it is not feasible in a conventional system or PTFBG operating with left light incidence. So far, the lowest value of nonreciprocal switching intensity reported in the context of nonlinear PTFBGs is 0.02 \cite{raja2019nonlinear}. We intend to reduce the value further by operating the PTFBG in the longer wavelength regime in the presence of chirping and modulation of the Kerr nonlinearity parameter as shown in Fig. \ref{fig5}(b). The switch-up intensity keeps on decreasing with any variation in the negative detuning parameter ($\delta < 0$). For a special case plotted in the inset of Fig. \ref{fig5}(b) with numerical values $C = -1$, $\gamma_K = -0.05$, $\gamma_{nl}$ = -1.5 and $\delta = -0.5$, the switch-up intensity value is measured to be 0.015 which is less than the lowest switching intensity reported by us previously \cite{raja2019nonlinear}. This once again proves that the quest to devise low power switches with the aid of newly engineered FBG is ever-expanding. As an end note to this particular section, we would like to stress that with a careful selection of other design parameters and the operating regime, the nonuniformities as well the perturbations in the grating can be turned into constructive ones (rather than being detrimental) in the context of realization of low power switches. It should be noted that the variation in the spectral span is effected as a consequence of the variation in the chirping and detuning parameters as shown in Figs. \ref{fig5}(c) and \ref{fig5}(d) under the condition that the light is launched from the rear end. This proves that the span over which the OB curves occur is independent of the light launching direction and it is dependent on chirping nonuniformity and the signal detuning parameter. 

	\subsection{Broken \texorpdfstring{$\mathcal{PT}$}--symmetric regime ($g>\kappa$)}
	\label{Sec:4b}
\begin{figure}[t]
	\centering	\includegraphics[width=0.5\linewidth]{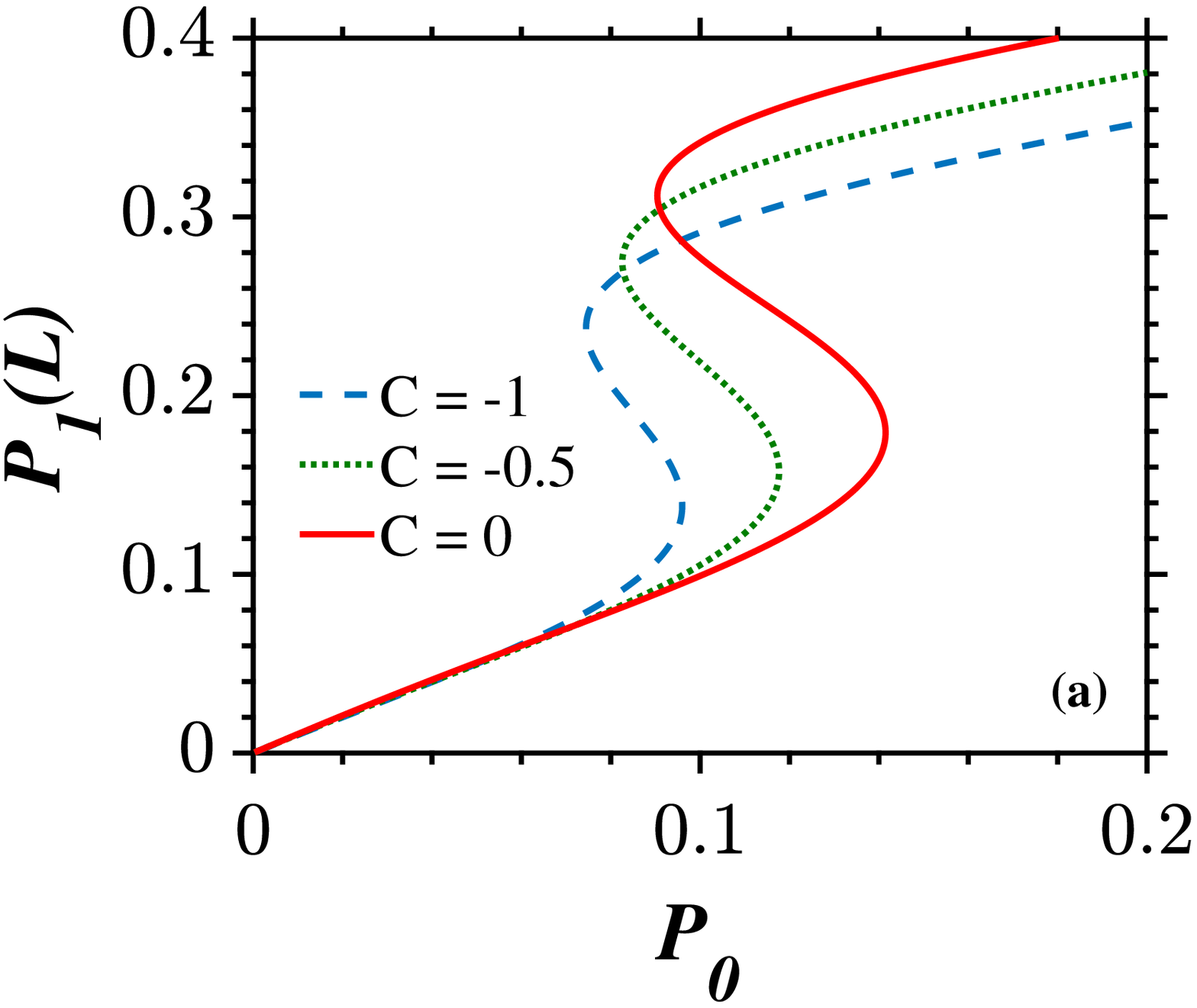}\includegraphics[width=0.5\linewidth]{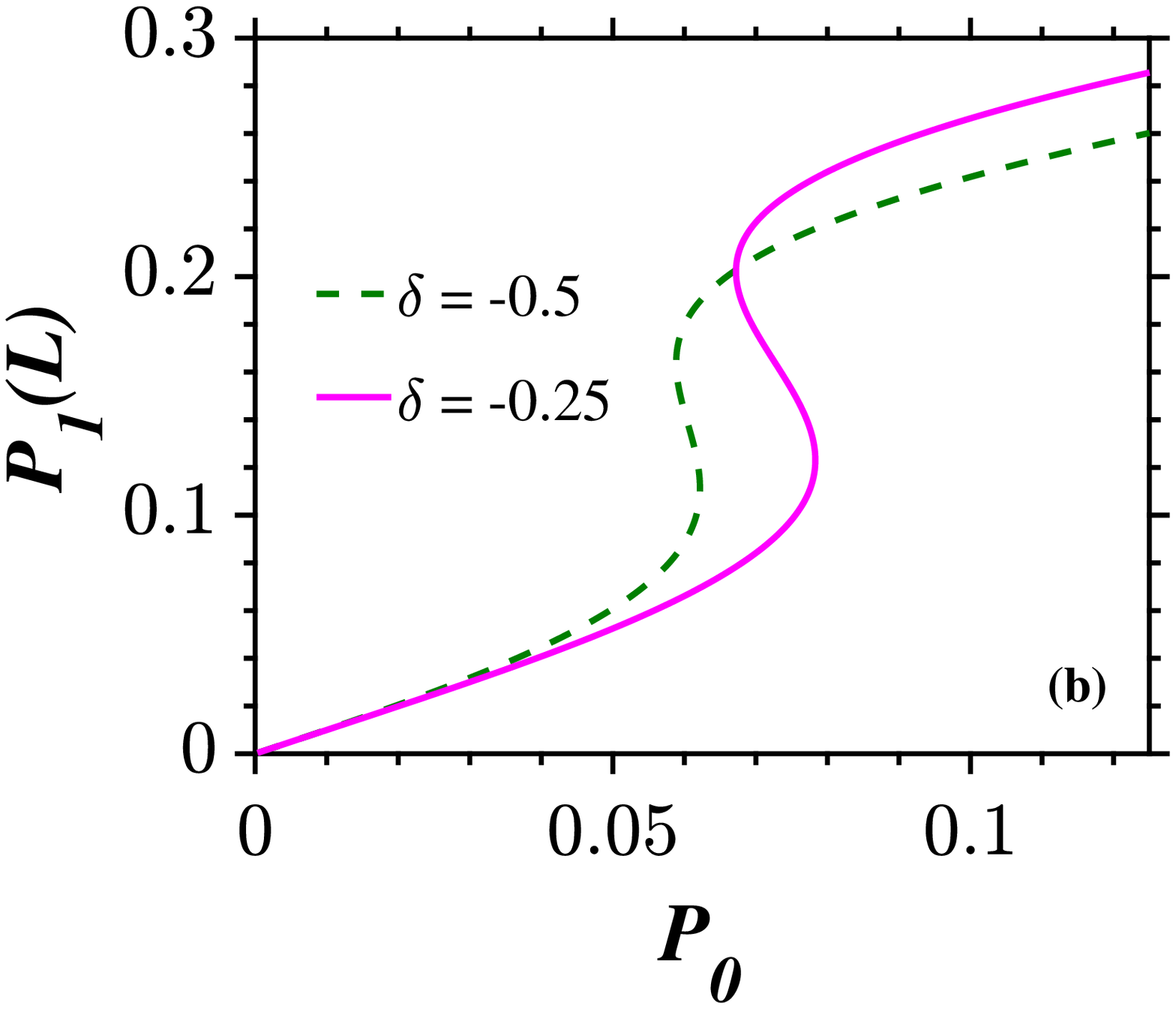}\\\includegraphics[width=0.5\linewidth]{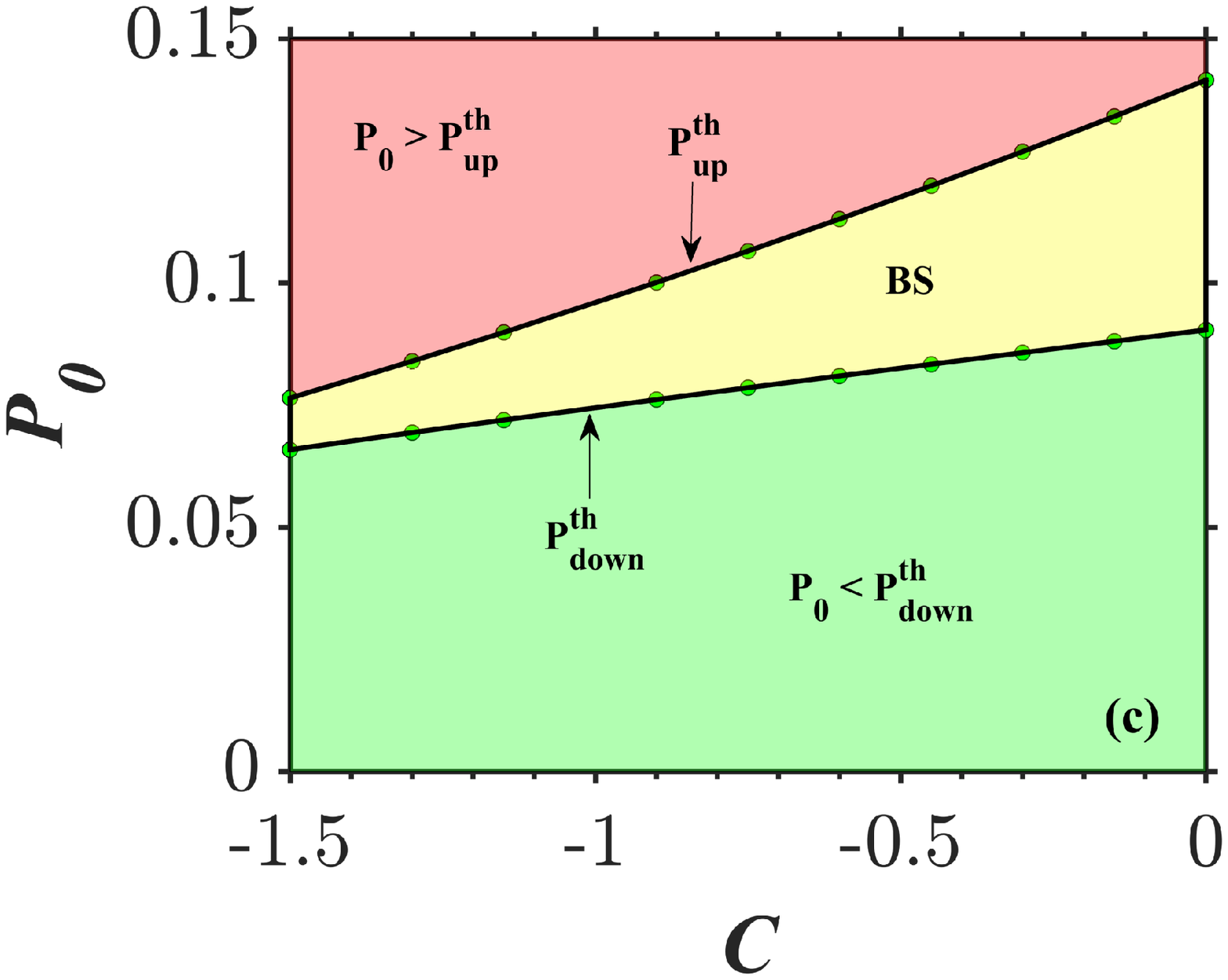}\includegraphics[width=0.5\linewidth]{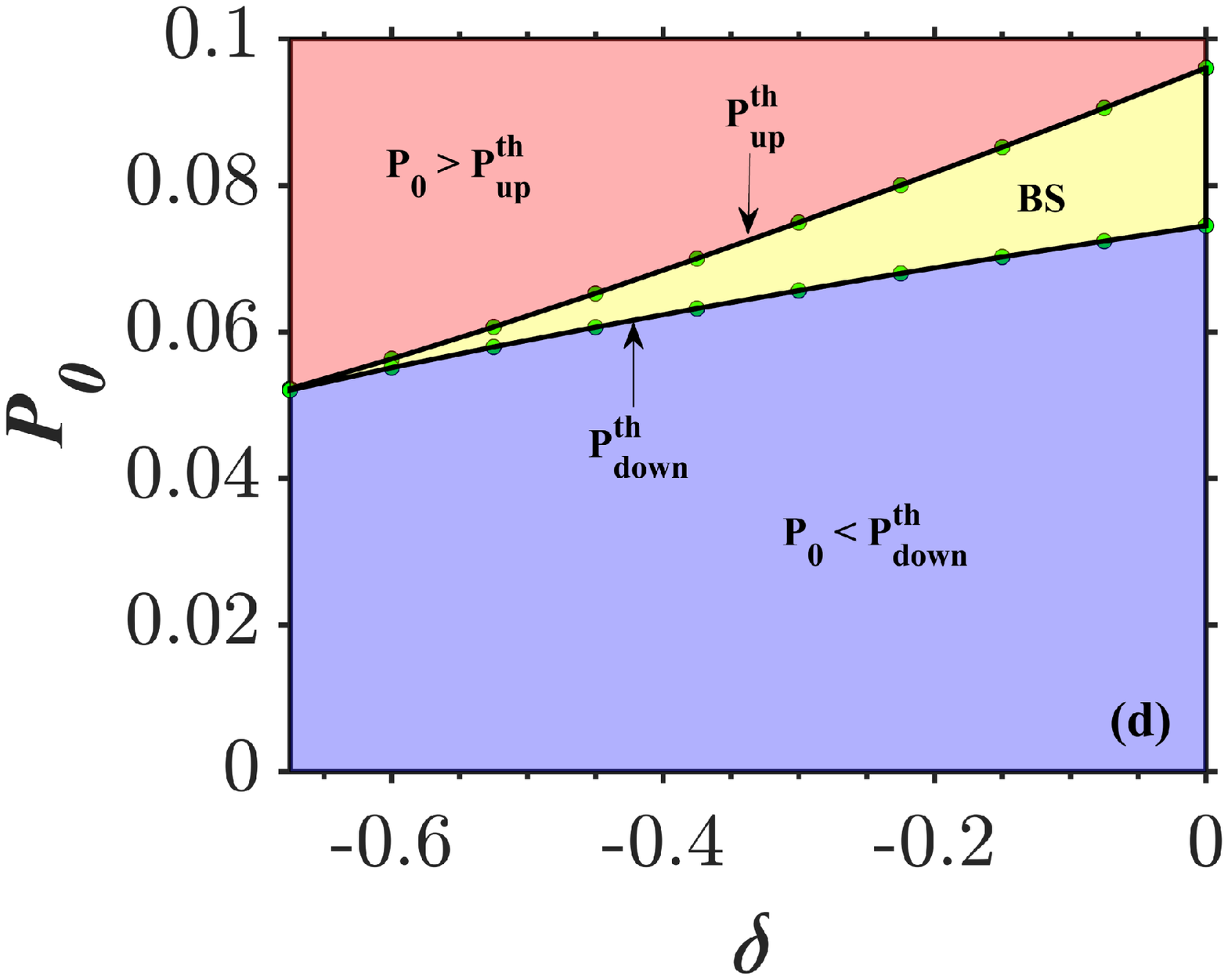}
	\caption{Plots showing the same dynamics with the same system parameters as in Fig. \ref{fig5} except that the light launching direction is right.}
	\label{fig6}
\end{figure}
In Sec. \ref{Sec:III}\ref{Sec:3b}, we disclosed the net effect of variation in the modulation of Kerr nonlinearity  ($\gamma_K$), chirping ($C$), and detuning ($\delta$) parameters on the ramp-like bistable states. Encouraged by this, we intend to study the existence of ramp-like stable states at lower powers under the adjustment in the light launching direction. The common features between Figs. \ref{fig4}(a) and \ref{fig6} are as follows: First, there exists a range of input intensities which is common to all the spectral components (before switching occurs). Second, the onset of switching is preceded by the occurrence of a ramp-like first stable state. Post the occurrence of switching, the second stable are well separated by significant amount of output intensities in Fig. \ref{fig6}(a). This proves that it is possible to achieve large spectral uniformity even at lower powers, thanks to the concept of nonreciprocal switching.  Figure \ref{fig6}(b) delineates the shift in the ramp-like bistable curves to the lower intensity side while operating in the longer wavelength regime. Operating the system in the broken $\mathcal{PT}$-symmetric regime under right light direction helps in altering the feature of spectral uniformity as shown in Figs. \ref{fig6}(c) and \ref{fig6}(d) similar to the conclusions drawn from Figs. \ref{fig4}(c) and \ref{fig4}(d)

	\section{Conclusions}
	\label{Sec:VI}
	Introducing modulation in the nonlinear refractive index in the form of the parameter $\gamma_K$ has proven to be a beneficial degree of freedom to realize low optical switches provided that it is of self-defocusing type alongside the self-defocusing cubic nonlinearity. The presence of gain and loss brings in nonreciprocal switching, low power steering dynamics, bistable curves with larger spectral nonuniformity. The sign of chirping and detuning plays a pivotal role in controlling the intensities at which steering occurs and the spectral span in which the OB phenomenon occurs. In the presence of high nonlinearity and modulation of the nonlinear index, manipulations in the nonuniformities like operation in the negative detuning regime accompanied by a simultaneous decrement in the spatial frequency of the  grating resulted in ultra-low power steering. The value of the critical intensity under these conditions is measured to be 0.015 (approximately) when the incident optical field is impelled into the CPTFBG from the rear end.  We hope that experimental researchers can bring the above theoretical results into reality in the near future.

	\section*{Acknowledgments}
	AG and ML acknowledge the support by DST-SERB for providing a Distinguished Fellowship (Grant No. SB/DF/04/2017) to ML in which AG was a Visiting Scientist. AG is now supported by the University Grants Commission (UGC), Government of India, through a Dr. D. S. Kothari Postdoctoral Fellowship (Grant No. F.4-2/2006 (BSR)/PH/19-20/0025).  
	
		\section*{Disclosures} 
	The authors declare no conflicts of interest.

\end{document}